\documentclass[%
 reprint,
 floatfix,
 amsmath,amssymb,
 aps,prx,
]{revtex4-2}
\usepackage{threeparttable}
\usepackage{xcolor}
\usepackage{varioref}
\usepackage{verbatim}
\usepackage[normalem]{ulem}
\usepackage{booktabs}
\usepackage{graphicx}
\usepackage{dcolumn}
\usepackage{bm}
\usepackage[hyperfootnotes=true,bookmarks=true,]{hyperref}
\usepackage{cleveref}
\hypersetup{final,		
			colorlinks=true,
			linkcolor=red,
            linktoc=page,
			anchorcolor=black,
			citecolor=red,
			urlcolor=blue}

\begin{document}

\title{Universalities in the Avalanche Dynamics of Novelties and Non-Novelties and some Notes on the Heaps law}

\author{Filippo Santoro}
\affiliation{CREF - Centro Ricerche Enrico Fermi, Via Panisperna, 89a, 00184 Roma RM}
\affiliation{Sapienza University of Rome, Physics Department, Piazzale Aldo Moro 5, 00185 Rome, Italy}

\author{Alberto Petri}
\affiliation{CNR - Institute of Complex Systems, Department of Physics, Piazzale Aldo Moro 5, 00185, Italy}
\affiliation{CREF - Centro Ricerche Enrico Fermi, Via Panisperna, 89a, 00184 Roma RM}

\author{Francesca Tria} 
\email{francesca.tria@uniroma1.it}
\affiliation{Sapienza University of Rome, Physics Department, Piazzale Aldo Moro 5, 00185 Rome, Italy}
\affiliation{Complexity Science Hub Vienna, Austria}
\affiliation{CREF - Centro Ricerche Enrico Fermi, Via Panisperna, 89a, 00184 Rome, Italy}

\begin{abstract}
Unprecedented events intertwine with the repetition of the past in natural phenomena and human activities.
Key statistical patterns, such as Heaps' and Taylor's laws and Zipf's law, have been identified as characterizing the dynamical processes that govern the emergence of novelties and the abundance of repeated elements. Observing these statistical regularities has been pivotal in motivating the search for modeling schemes that can explain them and clarify key mechanisms underlying the appearance of new elements and their subsequent recurrence. 
In this study, we analyze sequences of novel and non-novel elements, referred to as avalanches, in real-world systems. We show that avalanche statistics provide a complementary characterization of innovation dynamics, extending beyond the three fundamental laws mentioned above.
Although arising from collective dynamics, some systems behave as a single instance of a stochastic process. Others, such as natural language, exhibit features that we can only explain by a superposition of different dynamics. This distinction is not apparent when considering Heaps' law alone, while it clearly emerges in the avalanche statistics. 
By interpreting these empirical observations, we also advance the theoretical understanding of urn-based models that successfully reproduce the observed behaviors associated with Heaps', Zipf's, and Taylor's laws. We derive analytical expressions that accurately describe the probability distributions of avalanches and the Heaps law beyond its asymptotic regime.  
Building on these results, we derive a scaling relation that we show also holds in real-world systems, indicating a form of universality in the dynamics of novelty.

\end{abstract}

\maketitle

\section*{Introduction}
Innovations and novelties are increasingly recognized as valuable perspectives for gaining insights into systems related to human activities, social behaviors, and biological evolution. Diverse efforts have been devoted to understanding and explaining the mechanisms driving societal transformations and creativity, from economic growth~\cite{thurner2010schumpeterian,saracco2015innovation} to the building of collective knowledge~\cite{dankulov2015dynamics,iacopini2018network,iacopini2020interacting}, to the emergence of language structures~\cite{gerlach2013stochastic}.
We all recall how the emergence and diffusion of a new viral variant triggered novel societal habits and attitudes, which in turn impacted environmental  awareness~\cite{SHETH2020280,HAN2023933}. 
This cascade effect often marks less dramatic events: a conversation with a new collaborator can lead to the opening of a new research line, listening to a song by an unknown artist can spark interest in their entire body of work, or a single word dropped in conversation can initiate a new topic of discussion.
We immediately note that we sometimes mention genuine innovations and sometimes novelties -- the former referring to something new to the world on a global scale, and the latter to the adoption or experience of something new for an individual or a group on a local scale. In what follows, we will treat innovation and novelties on a common ground since their dynamics exhibit shared behaviors~\cite{tria2014dynamics}, even though they are distinct concepts.
Let us describe an innovation system as an ordered sequence of elements or events. This perspective encompasses a wide variety of different systems, from a sequence of genes sampled in a population to the sequence of songs listened to by users of a website, to the sequence of words that compose a book, to the sequence of contacts in social networks~\cite{cattuto2007heaps,tria2014dynamics,Ubaldi_2021}.
In this view, three fundamental laws hold: the Heaps~\cite{heaps1978information}, Zipf~\cite{zipf35}, and Taylor~\cite{taylor1961aggregation} laws.
The Heaps law describes the emergence rate of novelties or innovations and is related to the Zipf law, which describes the frequency distribution of the distinct elements in a system~\cite{lu2010zipf,Loreto2016}. 
The Taylor law~\cite{taylor1961aggregation} was initially introduced in ecology, relating the variance of a population density to its mean. A departure from a proportionality relation indicates a correlation in the distribution of individuals; specifically, variance growing more rapidly than the mean is a signal of mutual attraction among individuals in a natural population. 
In the framework of language~\cite{gerlach2014} and of innovation systems at large~\cite{tria2018zipf,tria2020taylor}, Taylor's law is applied to the relation between the variance and the mean of the number of innovations that emerged from the onset of the system until a given moment. It turns out that the variance grows approximately with the square of the mean, pointing to positive correlations in the emergence of novelties~\cite{tria2018zipf,tria2020taylor}.

Here, we explicitly address the correlation in the temporal occurrence of innovations from its statistical perspective by considering the occurrence of consecutive sequences of novelties and non-novelties, i.e., {\em avalanches} of novelties and non-novelties. 

Avalanches are referred to 
as phenomena that display randomly distributed events interspersed by inactivity or events of a different nature. They have received much attention because of their widespread presence in systems of a very different nature,  from the paradigmatic case of earthquakes \cite{Main1996,Bizzarri2019,Bizzarri2021},  to magnetic hysteresis \cite{Zapperi1998}, structural phase transitions \cite{Cannelli1993,Petri1994,Caldarelli1996}, plastic deformation \cite{Dimiduk2006}, granular shearing \cite{Baldassarri2006,Petri2008,Picaciamarra2012,Baldassarri2019,Petri2024}, 
internal sliding  \cite{Antonaglia2014}, as well as in neural activity \cite{de2006self,Lombardi2012,Nandi2022,scarpetta2023criticality}, and in the mobility of humans \cite{Chialvo2015} and of ants \cite{Gallotti2018}. 
Distinctive quantities of avalanches, such as size and duration,  are generally observed to display statistical distributions in the form of power laws in non-trivial systems. This behavior is often connected to the presence of some critical point \cite{Sethna2001,de2012activity,scarpetta2013neural,Pontuale2013,Petri2018,Zapperi2022} and of some form of scaling, i.e., in a given system, distributions related to different system parameters can collapse onto a single, universal,  invariant shape if expressed as a function of a suitable variable. This points to the presence of universal behaviors \cite{Laurson2013,de2006universality,Dearcangelis2016,Petri2018}. 
Moreover, it has been pointed out~\cite{kumar2020interevent} that systems that display similar avalanche distributions sometimes feature different inter-event time distributions, suggesting that the latter can
provide a complementary characterization that helps understand the process's dynamics and discriminate among different hypotheses and models. 

Here, we investigate the avalanche statistics of novelties and non-novelties in several innovation systems. We define an avalanche of novelties as a sequence of only consecutive novelties. Of particular interest are the intervals (or {\em inter-times}) between two successive novelties, which we also name avalanches of non-novelties, i.e., sequences of only consecutive non-novelties.  
While avalanches of novelties are very short, the lengths of the avalanches of non-novelties span different orders of magnitude, so that their distribution provides helpful insights into understanding the underlying dynamics and assessing the goodness of generative models.

We find that systems that behave similarly regarding the three fundamental laws discussed above—the Heaps, Zipf, and Taylor laws—exhibit different behaviours concerning the avalanche statistics of novelties and non-novelties.
In particular, among the considered systems, some exhibit a form of the inter-times distribution featuring an exponential-like decay. In contrast, others feature a fatter tail.
Further, the Corpus of English written texts displays a clear crossover between two power law regimes.

To gain insights into these results, we refer to a class of Polya's urn-based models proposed in~\cite{tria2014dynamics} that successfully account for the three fundamental laws mentioned above
and acts as reference models for various research in innovation systems and complex networks analysis~\cite{di2025dynamics,Ubaldi_2021,iacopini2020interacting}.
We will focus in particular on the urn model with triggering (UMT) and a specific version of it, exhibiting exchangeability~\cite{definetti,kingman1978random} and here dubbed UMT-E, that provides an urn characterisation of the two-parameter Poisson-Dirichlet process or Pitman-Yor process~\cite{pitman_1997,tria2020taylor} and was recently used in inference problems~\cite{tani2024inference}. 
The focus on the  UMT-E model is motivated by its analytical convenience. However, we show in the Supplementary Information (S8) that the same results can be obtained by referring to the general UMT  model, as well as to a generalization of it, accounting for semantic correlations, the urn model with semantic triggering (UMST)~\cite{tria2014dynamics}.

Through the lens of the UMT model, we find that the avalanche size statistics allow us to distinguish between processes that, although resulting from the contribution of many agents, can be described as a single collective dynamics and processes that require an explicit description in terms of a superposition of individual dynamics.
In particular, we find a perfect agreement between observations and model predictions for the datasets featuring a closed-to-exponential tail. At the same time, we argue that the fat-tail distributions are an output of collective dynamics.
 
 Moreover, we observe that, where a superposition of individual dynamics can account for the observed avalanche statistics, a single realization of the UMT can still reproduce it, though for different parameters than those that reproduce the Heaps law of the process. This observation could help understand the relation between the individual and the collective innovation rate.

As a support for the analysis, and as a value per se, we provide
 analytical expressions both for the Heaps law in the UMT model, beyond the asymptotic exponent, and for the avalanche size distribution of novelties and non-novelties.
 We firstly discuss predictions for the UMT-E, where a closed form for the average number of distinct elements as a function of the sequence length, expressing the Heaps law, can be written.
Further, we provide the predictions of the general non-exchangeable version of the UMT, where it is possible to write an inverse relation for the Heaps law, namely the average time corresponding to a given number of distinct elements. One can recover the asymptotic relation provided in~\cite{tria2014dynamics} by inverting this exact relation up to the leading order. In addition, we show that the inverse relation is sufficient to obtain the avalanche size distribution of novelties and non-novelties.
As a final result, we predict a scaling relation for the avalanche distributions from the UMT model and successfully apply it to the considered systems. 

We organize the paper as follows.
The Results section consists of three main parts. First, we report and comment on the results for the novelty and non-novelty avalanche size statistics in several systems featuring innovation. We contrast these results with the predictions of the UMT model. Second, we present analytical results for the main observables we consider, namely the number of distinct elements as a function of the sequence length, which expresses Heaps' law, and the probability distribution of avalanche lengths for both novelties and non-novelties. We finally derive the scaling relation for the avalanche distributions from the UMT-E model, and we test it in all the considered systems, as well as in the general UMT and the generalized UMST models. We find a striking agreement with the theoretical predictions in all the cases.
A Methods section follows, detailing the techniques adopted in the data analyses and numerical simulations. The Supplementary Information provides more details.

\section*{Results}

\subsection*{Avalanches of novelties and non-novelties in real data}

Innovation is an intermittent process,  for which real systems lie between the drive to explore the adjacent possible and the need to exploit the already known.
An ordered sequence can be divided into sub-sequences according to their nature, composed of new or not new elements, so avalanches of novelties and non-novelties can be defined.
 
\begin{figure}[h!]
	\centering
	\includegraphics[width =1\linewidth]{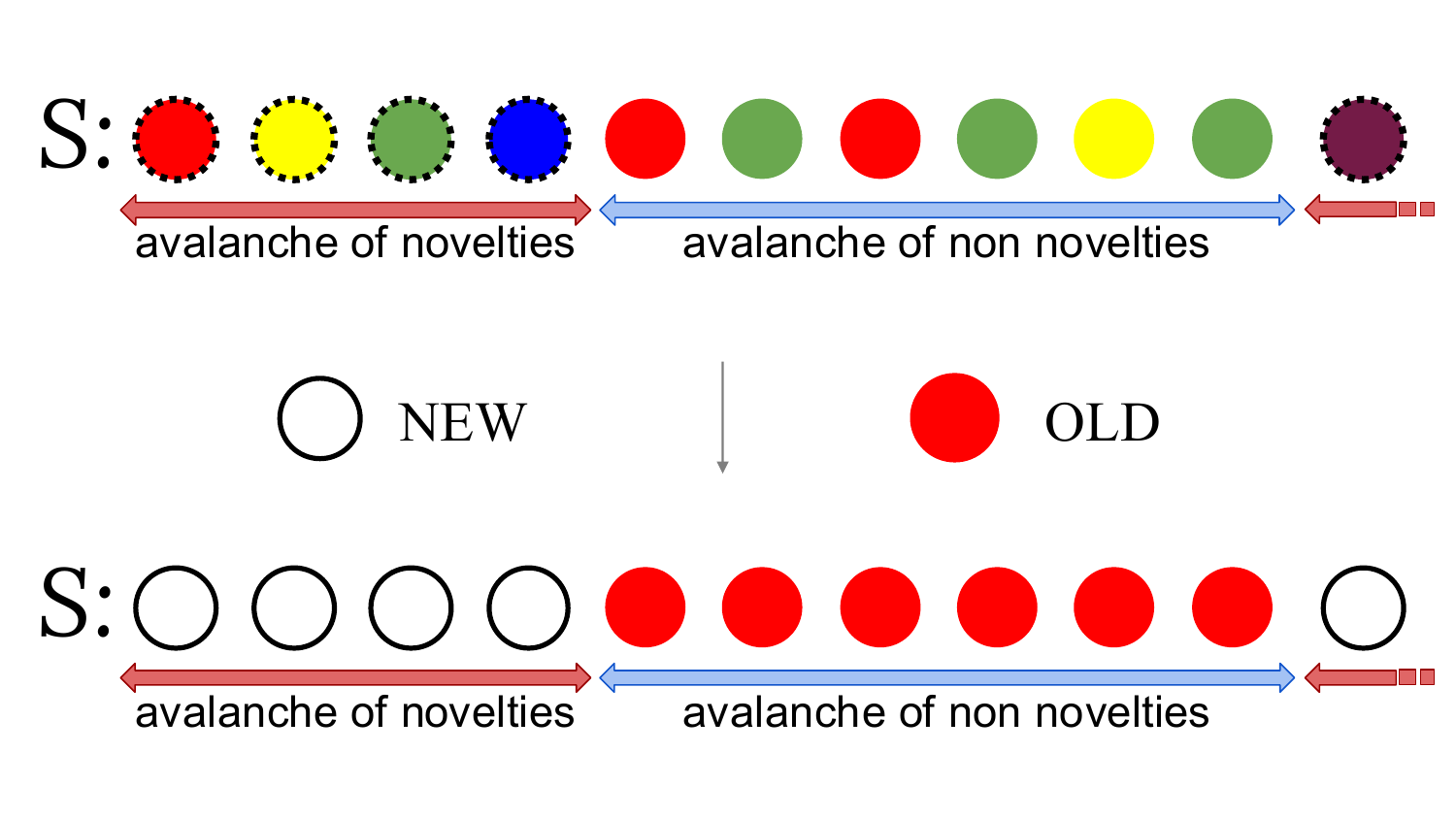}
	\caption{Cartoon of a sequence $S$ of events (flowing from left to right) where we highlight avalanches of novelties and avalanches of non-novelties. Top: coloured balls encircled with dotted lines represent novelties, i.e., elements that have never appeared in the sequence until that moment; coloured balls with no contour line represent non-novelties, i.e., elements already present.
Each color represents a distinct element that, depending on the context, can be: a word, a song, an edit, a hashtag, etc. Bottom: same as in the top, but we only characterize elements by being a novelty (white balls) or a  non-novelty (red balls).}
	\label{fig:avascheme}
\end{figure}

 An avalanche of novelties is a sub-sequence composed of elements new to the system, bounded on the left and right by two elements that are not new. An avalanche of non-novelties is a sub-sequence composed of elements not new to the system, bounded on the left and right by two new elements. Fig.~\ref{fig:avascheme} schematizes sub-sequences of consecutive novelties or non-novelties. 
 Avalanches of non-novelties can be understood as intervals between two novelties, and vice versa. In the following, we will use these expressions interchangeably.

We here analyze the size distribution of novelties and non-novelties avalanches in four databases related to web-mediated human activity  (LastFM, Github repositories, Twitter Hashtags,  Wikipedia) and one dataset related to non-web-mediated cultural production (The Gutenberg corpus of digitized English books). Moreover, in the light of the results in~\cite{Ubaldi_2021}, which show that the expansion into the adjacent possible can account for the human dynamics in social networks, we also analyse the avalanche statistics in the sequences of users' activity in the Twitter and Github datasets. All the systems considered share common statistical features expressed by the  Zipf, Heaps, and Taylor laws~\cite{tria2014dynamics,Monechi2017,tria2020taylor}.

We can formalize each dataset as a sequence $\mathcal{S}$ of elements $s_i$, $i=1,\dots,t_{\text{max}}$, where we can think of each index as a discrete time and where the values that the $s_i$ can take depend on the dataset: an element is a word in the case of books and Wikipedia pages,  an hashtag in the case of Twitter hashtags, a repository in the case of Github, a person in the Twitter and Github users datasets, a song in LastFM. 

Let us label each element $s_i$ with a label $x_i=0$ if it is a novelty, i.e., if its value did not appear in the sequence at times $t<i$, and $x_i=1$ otherwise. For instance, in the Gutenberg and Wikipedia datasets, a novelty is a word read for the first time by a reader who reads the sequences of pages or books in a given order set once and for all; in the sequences of users' activity in Twitter and Github, a novelty is a user who appears in the dataset for the first time; and so on.
We provide details on the datasets in the section Methods.

We call an avalanche of novelties of size (or length) $l$ a subsequence of $l$ contiguous elements with label $0$, preceded and followed by at least one element with label $1$; vice versa for an avalanche of non-novelties. 

\begin{figure*}[htbp!]
	\centering
	\includegraphics[width=\linewidth]{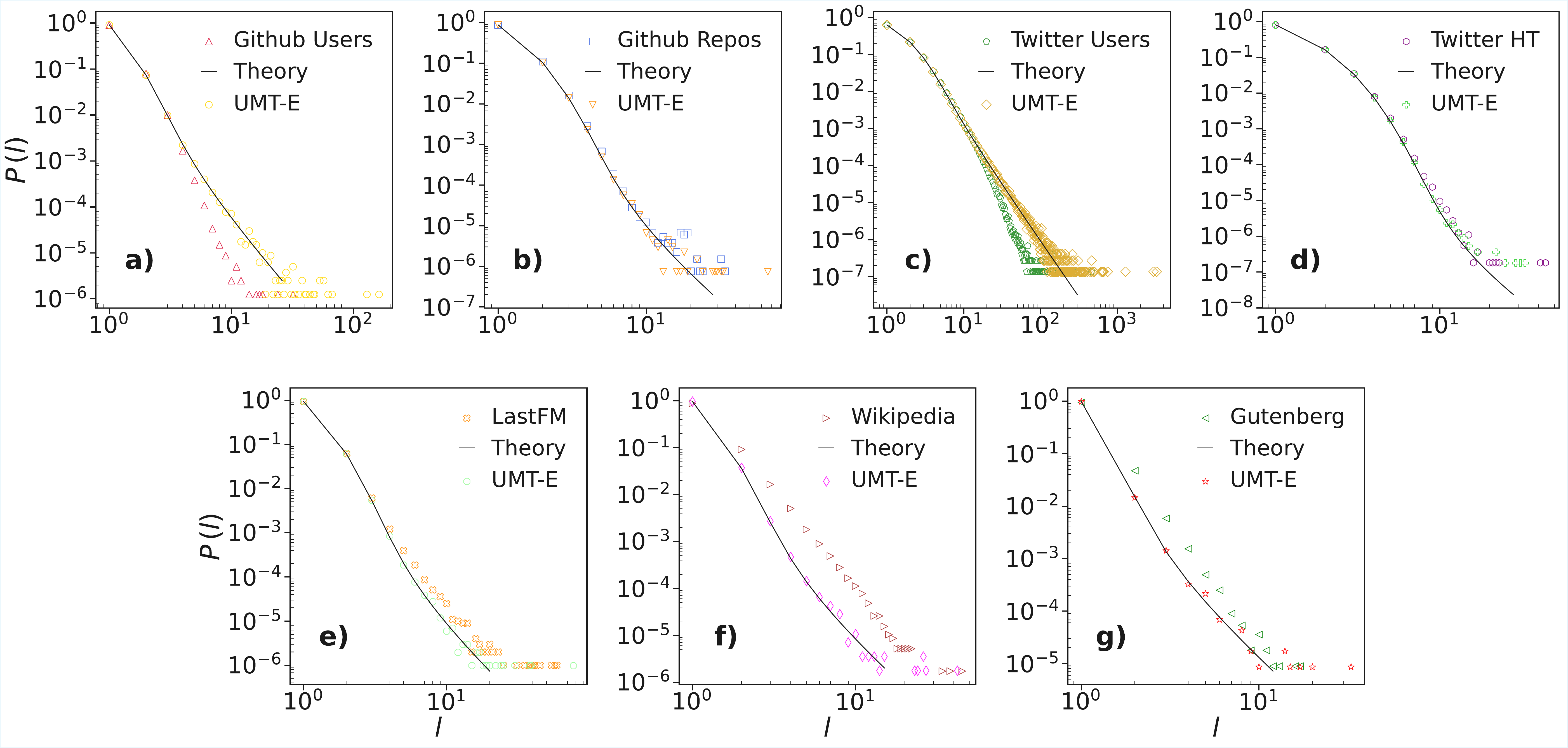}

	\caption{ \textbf{Avalanche size distribution for novelties.} We show results for the empirical datasets, contrasting them with predictions from the UMT-E model, as obtained through analytical computation and numerical simulations.
The parameters used for the UMT-E model predictions are those obtained from the fit of the Heaps law in the empirical datasets, and we report them in brackets for each dataset. \textbf{a)} Github Users $(\nu=4,\rho=7,N_{0}=45468)$. \textbf{b)} Github Repositories $(\nu=22,\rho=29,N_{0}=14640)$. \textbf{c)} Twitter Users $(\nu=16,\rho=29,N_{0}=42183043)$. \textbf{d)} Twitter Hashtags$(\nu=19,\rho=22,N_{0}=1995)$. \textbf{e)} LastFM$(\nu=28,\rho=41,N_{0}=27200)$. \textbf{f)} The Wikipedia Corpus $(\nu=31,\rho=50,N_{0}=25000)$.  \textbf{g)} the Gutenberg Corpus $(\nu=9,\rho=19,N_{0}=5890)$. We have truncated the sequences of both Gutenberg and Wikipedia at length $2 \cdot 10^{7}$.}
	\label{fig:panel_nov_full}
\end{figure*}

 As could be expected, the avalanches of novelties are small, spanning only an order of magnitude. We report the avalanche size distributions for the considered datasets and the corresponding results from the UMT-E model in Fig.~\ref{fig:panel_nov_full}.
 
\begin{figure*}[htbp!]
	\centering
	\includegraphics[width=\linewidth]{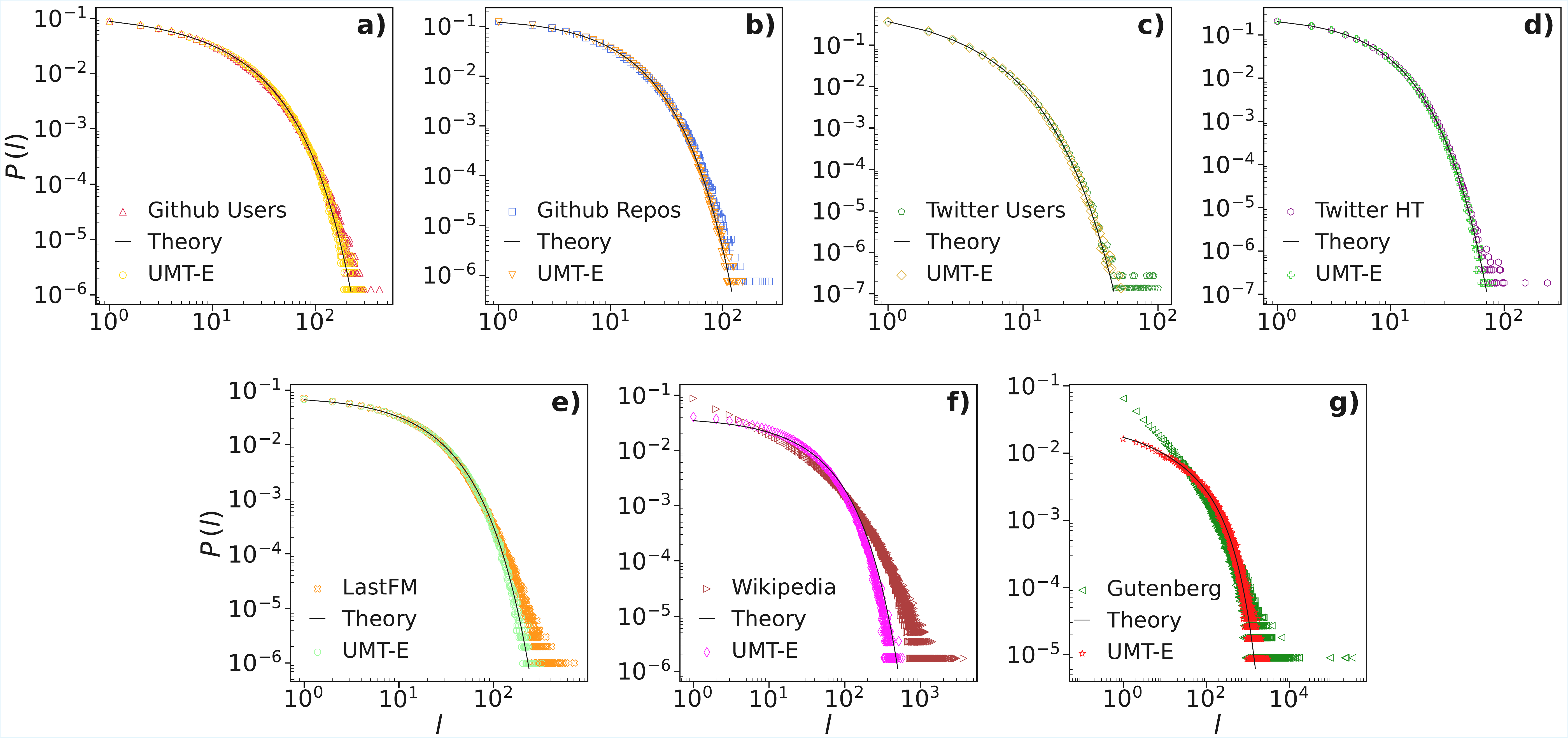}

	\caption{\textbf{Avalanche size distribution for non-novelties.} This plot is the analogue of Fig.~\ref{fig:panel_nov_full} for avalanches of non-novelties.}
	\label{fig:panel_non_nov_full}
\end{figure*}

 More interesting are the avalanches of non-novelties,
or equivalently, the inter-time distributions between two successive novelties, since they span several orders of magnitude.
We show them in Fig.~\ref{fig:panel_non_nov_full},
along with the theoretical prediction from the UMT-E model and the corresponding numerical simulations. 
To set the model's parameters, we fit the Heaps law by using the theoretical prediction discussed below (please refer to the fit in S1 of SI).
We use the obtained parameter values to generate sequences from the UMT-E and in the analytical estimate for the avalanche size statistics.
 
\begin{figure}[htbp!]
	\centering
	\includegraphics[width=\linewidth]{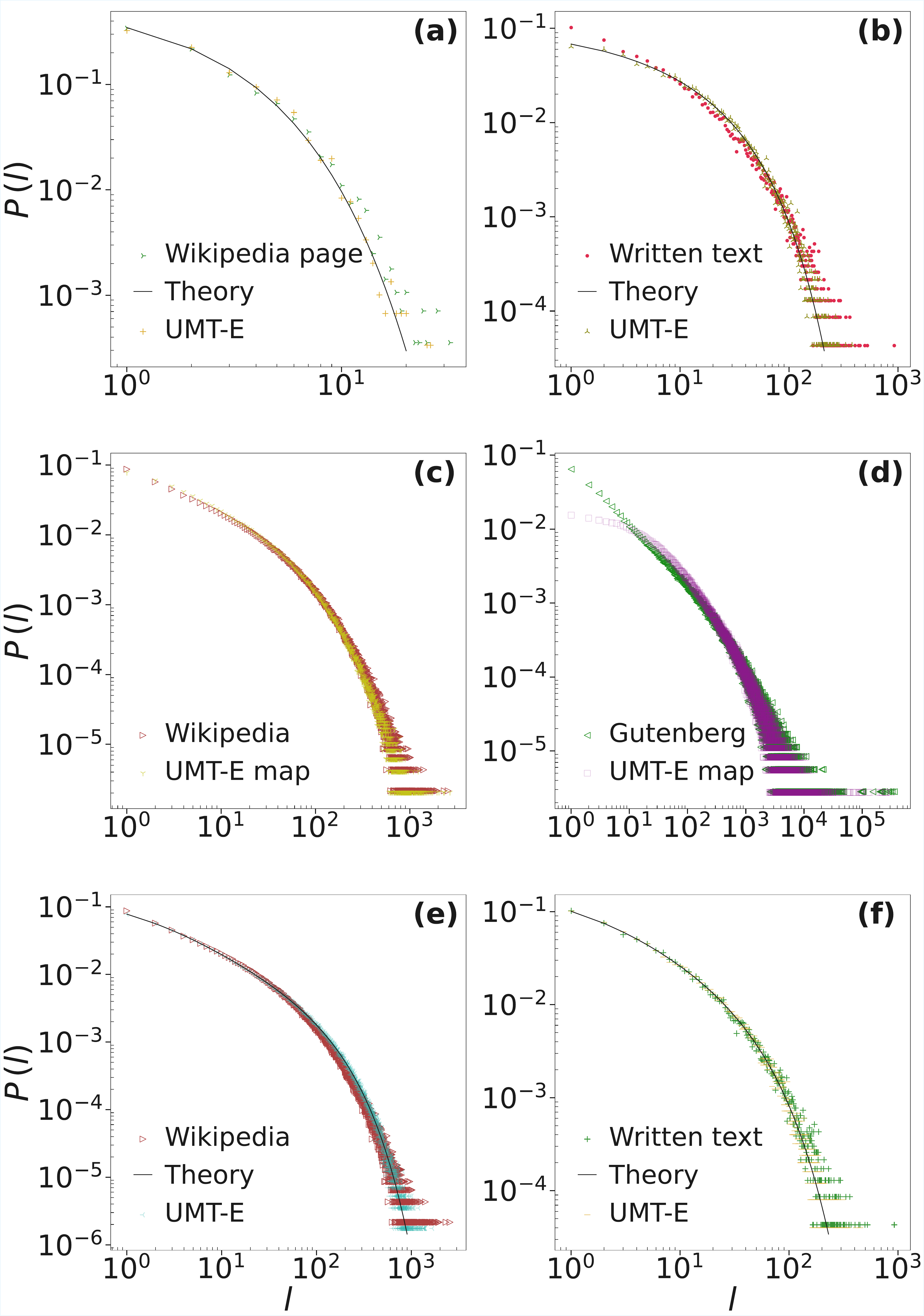}
	\caption{\textbf{Avalanche size distribution for non-novelties in the natural language corpora.} Top: We contrast results for the empirical datasets with predictions from the UMT-E model obtained through analytical computation and numerical simulations. The parameters
used for the UMT-E model predictions, reported in brackets, are those obtained from the fit of the Heaps law in the corresponding empirical data:  \textbf{a)} A single Wikipedia page $(\nu=4,\rho=7 , N_{0}=45468)$; \textbf{b)} a single book $(\nu=50,\rho=97, N_{0}=22912)$. Center: We contrast results for the empirical datasets, i.e., \textbf{c)} the Wikipedia corpus and  \textbf{d)} the Gutenberg corpus, with predictions from numerical simulations of multiple realizations of the UMT-E model, stacked one after the other, as described in the main text. In the case of Wikipedia, we have considered sequences with more than 500 words. Bottom: 
We contrast results for the empirical datasets with predictions from the UMT-E model obtained through analytical computation and numerical simulations. The parameters used for the UMT-E model predictions, reported in brackets, are those obtained from the fit of the avalanche size distribution for non-novelties in the corresponding empirical data: 
 \textbf{e)} The Wikipedia corpus $(\nu=2,\rho=9, N_{0}=400000)$; \textbf{f)} a single book $(\nu=23,\rho=73, N_{0}=127020)$.}
	\label{fig:panel_non_nov_wiki_and_gut}
\end{figure}

 We observe that the model perfectly reproduces the behavior displayed by the Github and Twitter datasets, related both to the users' production and activity (Fig.~\ref{fig:panel_non_nov_full}(a-d)).
 The inter-time distribution in LastFM (Fig.~\ref{fig:panel_non_nov_full}(e)) features a slightly larger tail than that reproduced by the model, while in the Gutenberg and the Wikipedia datasets it features a fat tail (Fig.~\ref{fig:panel_non_nov_full}(f-g)).
Although both referred to the written English language, the last two corpora  show a clear difference.
The inter-times distribution of the Gutenberg corpus displays a power law behavior with a double slope, which is absent in the Wikipedia corpus. To better understand the origin of this difference, we investigate the behavior of single books and single Wikipedia pages. We find that the UMT-E model perfectly reproduces the avalanche size statistics of single Wikipedia pages (Fig.~\ref{fig:panel_non_nov_wiki_and_gut}(a)). Conversely, single books exhibit a slight excess in short avalanche sizes and a slightly fatter tail than the corresponding sequences from the model, but thinner than for the whole Corpus (Fig.~\ref{fig:panel_non_nov_wiki_and_gut}(b). Please refer to S3 of SI for further examples).
We hypothesize, therefore, that the fat tail in the inter-times distribution in the Wikipedia and Gutenberg corpora derives from the convolution of different dynamics of single Wikipedia pages or single books, respectively. To test this hypothesis, we devised an experiment in which we separately fit the Heaps law for all the books in the Gutenberg corpus. We then generate sequences from the UMT-E model with the fitted parameters and stack them one after the other to form a unique sequence, as with real data.
We performed the same experiment on the Wikipedia pages. This procedure requires a suitable mapping between elements in the generated sequences and words (we leave a complete hierarchical model for further research, beyond the scope of this work). Here, we use a direct mapping by identifying the most popular element in the generated sequence with the most popular word in the corresponding book or Wikipedia page, the second most popular element with the second most popular word, and so on (please refer to the Method section for details). We show the results of these experiments in Fig.~\ref{fig:panel_non_nov_wiki_and_gut}(c-d), respectively for the Wikipedia and the Gutenberg corpus. We can see that the synthetic data perfectly reproduce the fat tail of the distributions of both datasets. The power law behaviour at short lengths observed in the Gutenberg corpus is, on the contrary, not explained by the bare UMT model.

As a last remark, we note that the inter-time distributions for the whole Wikipedia corpus and for single books can be well reproduced by a single UMT-E process with suitable parameter values, if we relax the constraint that the model should also fit the Heaps law (Fig.~\ref{fig:panel_non_nov_wiki_and_gut}(e-f)). This analogy between the whole Wikipedia corpus and single books suggests that the coexistence of different topics within a single book can play a key role in shaping the avalanche statistics in the literary Corpus.

In the next section, we derive the theoretical prediction for the Heaps law and the avalanche statistics for the UMT-E and the general UMT model.

\subsection*{Predictions from the urn model with triggering}

In the urn-based models proposed in~\cite{tria2014dynamics}, sequences of events are constructed through repeated extraction from an urn containing all the possible outcomes. The content of the urn changes along the evolution of the system, always allowing the entrance of brand new elements. Depending on the context, elements in the urn represent: words in a book or in a Wikipedia page, songs listened to on a music website, edits in collaborative software projects, hashtags on Twitter, and so on. The sequence of elements drawn from the urn represents the realized history. 

\subsubsection*{The model}\label{sec:UMT}

The urn model with triggering (UMT) is a generative model introduced in \cite{tria2014dynamics}. 
It is inspired by the adjacent possible expansion~\cite{kauffman_2000} as the framework for the emergence of novelties, and predicts the power laws for the Heaps, Zipf, and Taylor laws~\cite{tria2014dynamics,tria2018zipf,tria2020taylor}.

\begin{figure*}[htbp!]
	\centering
	\includegraphics[width=\linewidth]{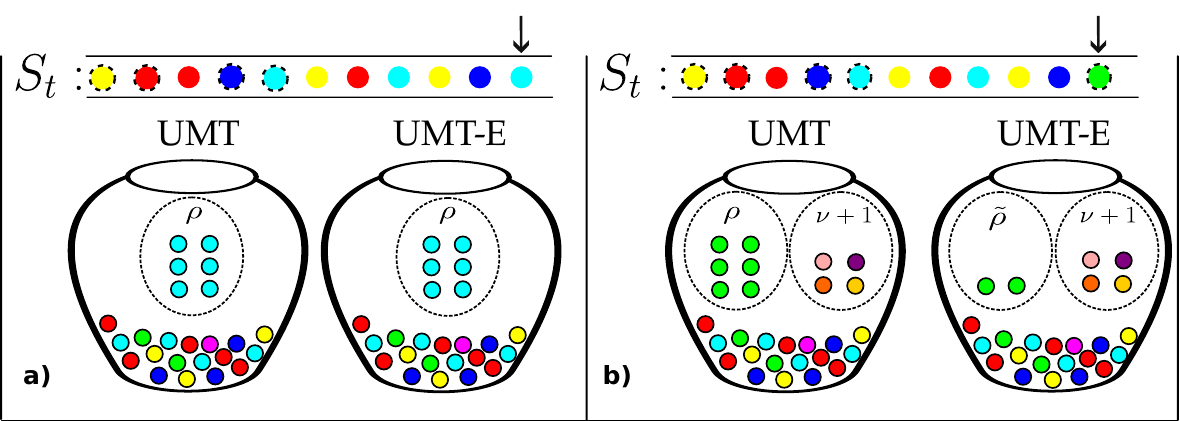}
	\caption{\textbf{Cartoon of the UMT process and its exchangeable version, UMT-E.}
    \textbf{a)} 
 The last extracted element (the cyan ball) has already appeared in the sequence, i.e, is a non-novelty. Only the reinforcement process occurs: we insert $\rho$ copies of the cyan ball into the urn in both models. 
    \textbf{b)}  The last extracted element (the lime ball) has never appeared in the sequence, i.e, it is a novelty. In this case, both reinforcement and triggering processes occur. The triggering process is identical for both models: we insert $\nu+1$ new distinct elements, whose colors were not present in the urn, into it. Conversely, the reinforcement process differentiates the two models: in UMT, we introduce $\rho$ copies of the last appended ball, while in UMT-E, we only introduce $\tilde{\rho}=\rho-(\nu+1)$ copies of it. Note that in the UMT-E model, we introduce the same number of balls in the urn at each step, independently of the last extracted ball.}
	\label{fig:explanation_umt_umte}
\end{figure*}

The process starts with an urn containing $N_{0}$ balls of different colors. A realization is obtained through the sequential random extraction of balls from the urn. Each time a ball is extracted, it is appended to the sequence $S$ and its color is reinforced, putting in the urn $\rho$ copies of it. If, in addition, the color is a novelty, that is, it appears for the first time in the sequence, then also
$\nu +1$ different balls of brand new colors are added to the urn (\textit{triggered innovation} process).
Let us now consider a slightly more general version of the model by modifying the reinforcing parameter $\rho$ for novelties. In particular, one can use the reinforcing parameter $\rho$
if the extracted element is a non-novelty, and a reinforcing parameter $\Tilde{\rho}$ if the extracted element is a novelty. This change does not impact the asymptotic exponents of the Heaps, Zipf, and Taylor laws~\cite{tria2014dynamics}.
When $\rho=\Tilde{\rho}$, we end up in the original UMT model. By fixing instead $\Tilde{\rho}=\rho-(\nu+1)$, in the case $\nu<\rho$, so that the number of balls inserted in the urn is constant at each step, independently on the drawn of a new or old element, the model becomes exchangeable, meaning that the probability of any generated sequence depends only on the numbers $n_i$ of each color $i$ in the sequence (and independently of the identity of the color $i$) and not on the order in which they appear. The exchangeable version of the model,  UMT-E, was shown~\cite{tria2018zipf,tria2020taylor} to be an urn representation of the two-parameter Poisson-Dirichlet process.
In Fig.~\ref{fig:explanation_umt_umte} we highlight the difference between the UMT and the UMT-E models.

For any value of $\Tilde{\rho}\geq 0$, the model predicts asymptotic power laws for the Heaps and Zipf laws,
of the form respectively $D(t)\simeq t^{\beta}$ and $f(R)\simeq R^{-\alpha}$, where $D(t)$ is the average number of distinct elements in a sequence of length $t$ and $f(R)$ is the fraction of elements of rank $R$, where elements are ordered according to their frequency, from the most to the least frequent. The exponents  depend on the relative importance of innovation versus reinforcement, through the ratio $\frac{\nu}{\rho}$, and read $\alpha = \frac{\rho}{\nu}$, and 
 $\beta = \frac{\nu}{\rho}$ when $\nu < \rho$, $\beta=1$ when $\nu > \rho$~\cite{tria2014dynamics}. Furthermore, the model also
 predicts the Taylor law with a relation $\sigma \simeq \mu^\gamma$, where $\gamma=1$ when $\nu < \rho$,  as approximately observed in real data, and spanning all the exponents between $1$ and $0.5$ when $\nu > \rho$, the value $0.5$ characterizing processes where innovation enter one after the other independently~\cite{tria2018zipf,tria2020taylor}.

The asymptotic exponents are enough to characterize the Heaps', Zipf, and Taylor laws. However, to provide an analytical expression for the avalanche distribution of novelties and non-novelties, we need to step further and seek an expression of $D(t)$ that is also valid at small time $t$. The initial number of balls $N_0$ is a crucial parameter for the exact form of $D(t)$ and for the shape of the avalanche statistics (see Fig.~\ref{fig:avalanches_umt_theo}).

\subsubsection{The Heaps law in the UMT-E}
\begin{figure}[htbp!]
	\centering
	\includegraphics[width=\linewidth]{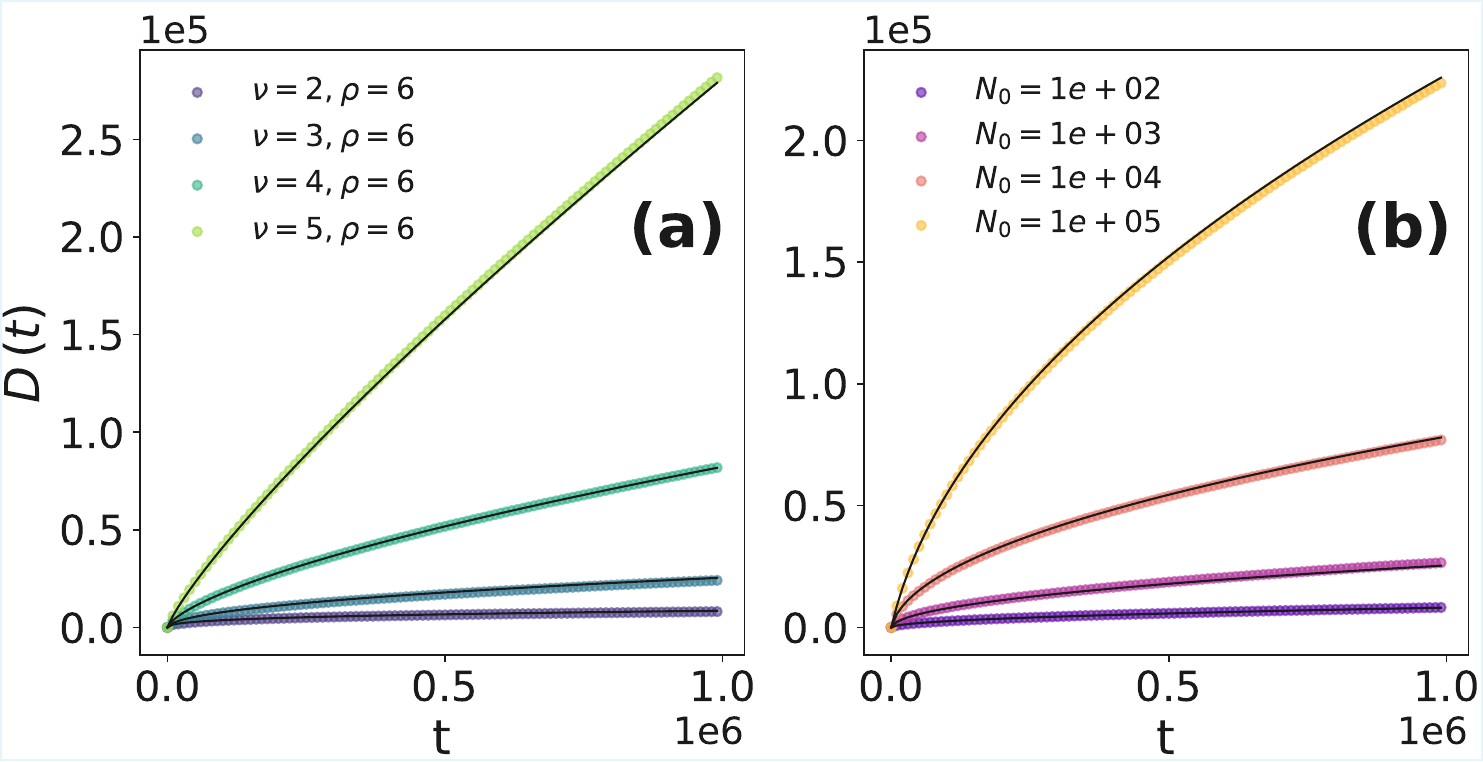}
	\caption{\textbf{The Heaps law in the UMT-E.} We show results for the Heaps law from numerical simulations of the UMT-E, along with the corresponding analytical predictions. \textbf{a)} We fix $N_{0}=1000 $ and $\rho=6$, and we consider different ratios $\frac{\nu}{\rho}$ by moving $\nu$ in the integer range $[2,5]$. \textbf{b)} We fix $\nu=3 $, $\rho=6$, and $N_{0}$ runs across different orders of magnitude from $10^{2}$ to $10^{5}$. All the sequences generated from numerical simulations have length $10^{6}$. }
	\label{fig:heaps_law_umt_e}
\end{figure}

The exchangeable version of the model (UMT-E) allows for an explicit expression of the average number of distinct elements $D(t)$ as a function of the total number of elements $t$.
We can solve the continuous-time differential equation.
\begin{equation}\label{eq:1}
\frac{dD(t)}{dt} = p_{\text{new}}(t) = \frac{N_0 + \nu D(t)}{N_0 + \rho t}
\end{equation}
with initial conditions $D(t=0)=0$. We obtain:
\begin{eqnarray}\label{eq:Dt_UMTE}
D(t) &= &\frac{N_0}{\nu} \left(1 + \frac{\rho t}{N_0} \right)^{\frac{\nu}{\rho}} - \frac{N_0}{\nu} = \nonumber\\
&=&\frac{\theta}{\alpha} \left(1 + \frac{ t}{\theta} \right)^{\alpha} - \frac{\theta}{\alpha}
\end{eqnarray}
\noindent
where in the last expression we set the identities $\frac{N_0}{\rho} \equiv \theta $ and $ \frac{\nu}{\rho}\equiv \alpha$ to stress the dependence of the exchangeable UMT-E model on only two parameters and make more evident the connection with the two-parameter Poisson-Dirichlet process, as discussed in~\cite{tria2018zipf,tria2020taylor}.
We note that the result in Eq.~(\ref{eq:Dt_UMTE}) could be derived, less straightforwardly, by taking the long sequence limit, from the exact solution of the two-parameter Poisson-Dirichlet process~\cite{pitman1996combinatorial}, or also from the results on triangular urn models in~\cite{zhang2015explicit}. In Fig.~\ref{fig:heaps_law_umt_e}, we show the theoretical result in Eq.~(\ref{eq:Dt_UMTE}) contrasted with results from simulations for different values of the ratios $ \frac{\nu}{\rho}$ (Fig.~\ref{fig:heaps_law_umt_e} (a)) and $\frac{N_0}{\rho}$ (Fig.~\ref{fig:heaps_law_umt_e} (b)).
We further report in the Supplementary Information (S5) the solution of Eq.~(\ref{eq:1}) with initial conditions $D(t=1)=1$, which more closely approximates the empirical curve when $N_0 \simeq 1$. 

 \subsubsection{The Heaps law in the general UMT}
We here provide the inverse relation $t(D)$ for the general UMT, with any definition of $\Tilde{\rho}\geq 0$. 
This relation cannot be inverted to provide an exact expression of $D(t)$ in the general case. We provide the expression for the leading term of $D(t)$, obtained by inverting $t(D)$, and discuss its relation to the asymptotic result in~\cite{tria2014dynamics}. 

\begin{figure}[htbp!]
	\centering
	\includegraphics[width=0.8\linewidth]{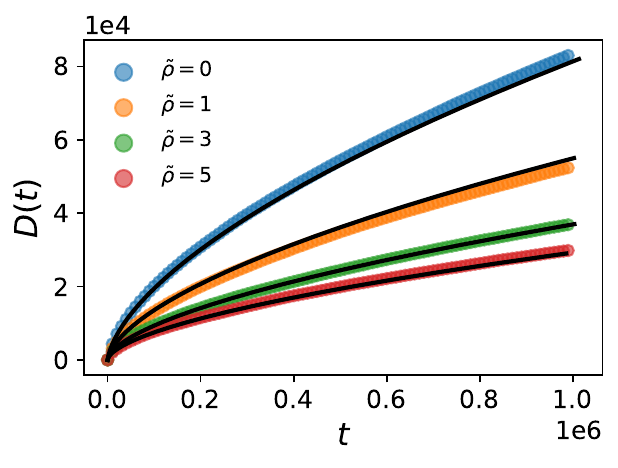}
	\caption{\textbf{The Heaps law in the general UMT.} We show results for the Heaps law from numerical simulations of the UMT, along with the corresponding analytical predictions. We set the parameters values to $(\nu=3,\rho=5$, and $N_{0}=1000)$, and we vary $\tilde{\rho}$ in the  range $[0,1,3,5]$. Continuous lines represent the theoretical predictions as discussed in the main text. All the sequences generated from numerical simulations have length $10^{6}$.}
	\label{fig:heaps_general}
\end{figure}

Let us then consider the general UMT model with $\Tilde{\rho}\geq 0$. The continuous time equation for $D(t)$ reads:
\begin{equation}\label{eq:heaps_generalUMT}
\frac{dD(t)}{dt} = p_{\text{new}}(t) = \frac{N_0 + \nu D(t)}{N_0 + \rho t + a D(t)}
\end{equation}
with $a \equiv \Tilde{\rho} + \nu+1 -\rho$, and for $a=0$ we recover the UMT-E, while for $a=\nu+1$ we recover the original UMT model.
We solve instead the inverse equation:
\begin{equation}
\frac{dt(D)}{dD} = \frac{N_0 + \rho t (D)+ a D}{N_0 + \nu D}
\end{equation}
that we can write as
\begin{equation}
\frac{dt(D)}{dD} - \frac{\rho }{N_0 + \nu D} t (D)=  \frac{N_0 + a D}{N_0 + \nu D}
\end{equation}
putting in evidence its nature as a non-homogeneous first-order linear differential equation.
Solving by standard methods and imposing the initial condition $t(0)=0$, we obtain the solution:
\begin{eqnarray}\label{eq:t_D_result}
t(D)&= &\frac{N_{0}}{\rho} \left(1 + \frac{a}{\rho-\nu} \right) \left(\frac{N_{0}+\nu D}{N_{0}} \right)^{\frac{\rho}{\nu}}  + \\
&- &a\left(\frac{N_{0}+\nu D}{\nu(\rho-\nu)} \right) -\frac{N_{0}}{\rho} \left( \frac{\nu-a}{\nu} \right) \nonumber
\end{eqnarray}
\noindent We show in Fig.~\ref{fig:heaps_general} the theoretical result in Eq.~(\ref{eq:t_D_result}) contrasted with simulated results of the model for different values of $\Tilde{\rho}\geq 0$,  finding perfect agreement. 

We also provide an approximate expression for $D(t)$, obtained by inverting $t(D)$ and retaining the leading-order term in $t$:
\begin{equation}\label{eq:t_D_approxresult}
D(t) \approx
\frac{N_0}{\nu}
\left(
\frac{\rho-\nu}{\rho-\nu+a}
\cdot
\frac{\rho t}{N_0}
\right)^{\frac{\nu}{\rho}}
\end{equation}
We note that the expression in  Eq.~(\ref{eq:t_D_approxresult}) agrees with that presented in~\cite{tria2014dynamics}  up to a multiplicative factor.
The asymptotic expression derived in~\cite{tria2014dynamics} was obtained by 
neglecting the parameter $N_0$, which represents the initial number of balls in the urn, and therefore cannot account for its effect. 
This parameter, however, turns out to play a crucial role in shaping the avalanche size statistics (please refer to  Fig.~\ref{fig:avalanches_umt_theo}). 
 In the Supplementary Information (S5), we compare the predictions for the Heaps law as well as for the avalanche size statistics obtained using the approximate expression in Eq.~(\ref{eq:t_D_approxresult}) with those obtained from the asymptotic solution in~\cite{tria2014dynamics}. 
We further report in the Supplementary Information (S5) the counterpart of Eq.~(\ref{eq:t_D_approxresult}) obtained by imposing the initial conditions $D(t=1)=1$, showing that it more closely approximates the empirical curve when $N_0 \simeq 1$.

\subsubsection{Avalanches statistics in the UMT-E}

Let us first focus on the probability distribution of the non-novelties avalanche size, and we then derive the corresponding probability for the novelties in an analogous way. 
Let us label as above $x_i=0$ if the element is a novelty, and $x_i=1$ if it is not a novelty (it is a non-novelty). 
We call an avalanche of non-novelties of size (or length) $l$ a subsequence of $l$ contiguous elements with label $1$, preceded and followed by at least one element with label $0$.

\begin{figure}[htbp!]
	\centering
	\includegraphics[width=\linewidth]{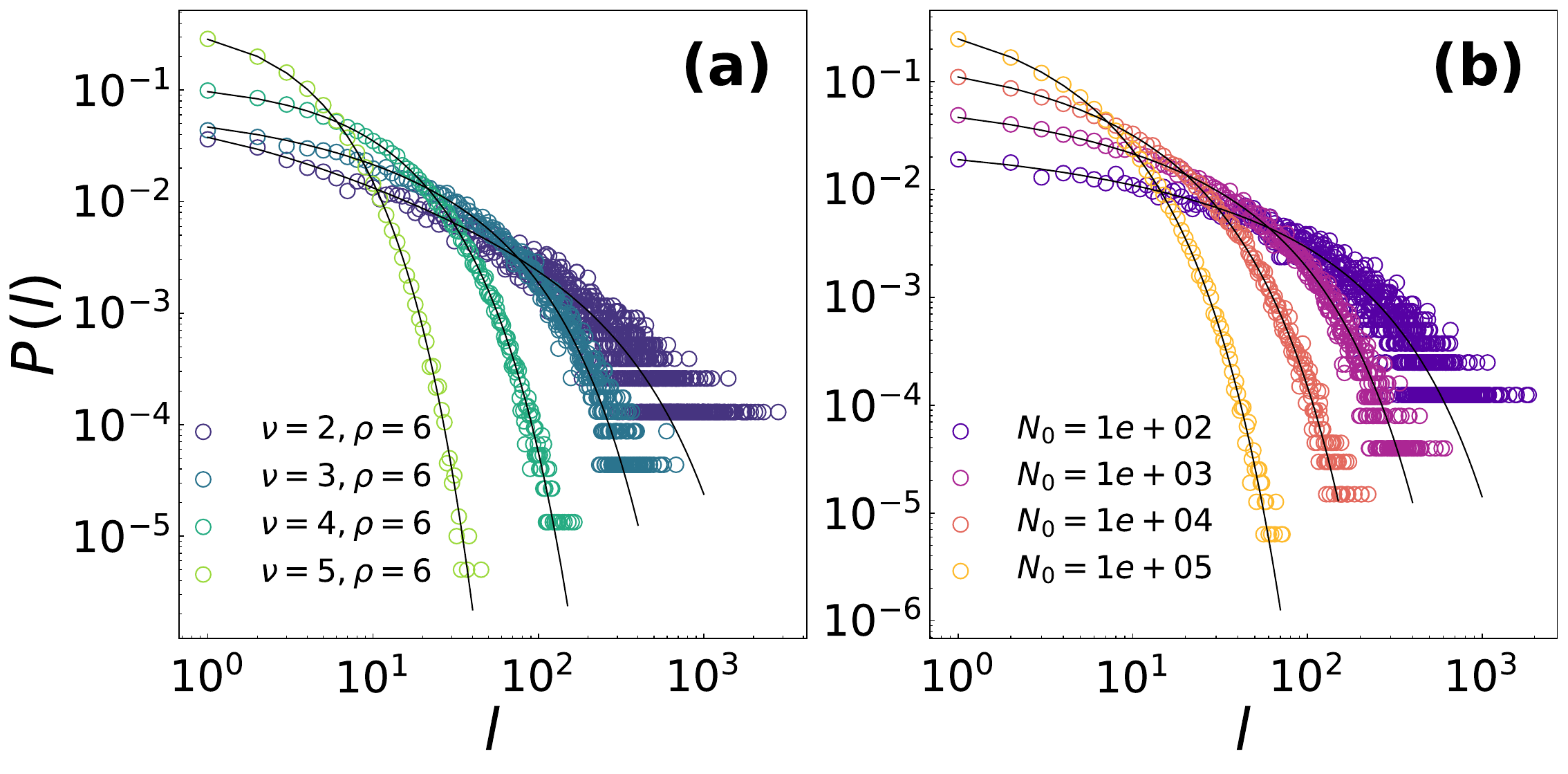}
	\caption{\textbf{The Avalanche size distribution for non-novelties in the UMT-E.} We show results from numerical simulations of the UMT-E, along with the corresponding analytical predictions. \textbf{a)} We fix $N_{0}=1000 $ and $\rho=6$, and we consider different ratios $\frac{\nu}{\rho}$ by moving $\nu$ in the integer range $[2,5]$. \textbf{b)} We fix $\nu=3 $,  $\rho=6$, and $N_{0}$ runs across different orders of magnitude from $10^{2}$ to $10^{5}$. All the sequences generated from numerical simulations have length $10^{6}$.}
	\label{fig:avalanches_umt_theo}
\end{figure}

We start by noting that the probability of finding that subsequence depends on the time at which we seek it, so we need to define a time-dependent probability
$p_t(x_{t-1},x_t,x_{t+1}, \dots, x_{t+l-1}, x_{t+l})$ such that $x_{t-1}=x_{t+l}=0$ and  $x_t,x_{t+1}, \dots, x_{t+l-1}=1$.
We can write:
\begin{eqnarray}\label{eq:pt}
&&p_t(x_{t-1},x_t,x_{t+1}, \dots, x_{t+l-1}, x_{t+l}) =\\
&&=p_{new}(t-1) p_{new}(t+l)\prod_{i=0}^{l-1}(1-p_{new}(i)) \simeq\nonumber\\
&& \simeq p_{new}(t)^2 ( 1-p_{new}(t))^l  \nonumber
\end{eqnarray}
where $p_{new}(t)$ is the probability of finding a new element at time $t$.
We then have to sum up over all the times from zero to $t_{\text{max}}$, where we can use the sequence length as $t_{\text{max}}$~\footnote{We should stop the integration at time $t=t_{\max}-l$, but neglecting this detail corresponds to the same order of approximation as that made in Eq.~(\ref{eq:pt}) and does not affect the results}. 
Furthermore, the non-novelties avalanche size distribution
is obtained by conditioning the above probability on the actual finding of a non-novelty avalanche, that is, on the probability of finding the pair \{{\it novelty, non-novelty}\}. The latter can be expressed by $p_t(0,1) = p_{new}(t-1) (1-p_{new}(t))\simeq p_{new}(t) (1-p_{new}(t))$.
By using Bayes' theorem, we finally obtain the non-novelties avalanche size
distribution:
\begin{equation}\label{eq:pl}
P_{\text{non-nov}}(l)=\frac{ \int_0^{t_{\text{max}}} p_{\text{new}}(t)^2 ( 1-p_{\text{new}}(t))^l  dt }{\int_0^{t_{\text{max}}} p_{\text{new}}(t) ( 1-p_{\text{new}}(t)) dt}.
\end{equation}
By substituting, from~Eqs.(\ref{eq:1}) and~(\ref{eq:Dt_UMTE}):
\begin{equation}
p_{\text{new}}(t) = \left(1 + \frac{\rho t}{N_0} \right)^{\frac{\nu}{\rho} - 1}.
\end{equation}
We can thus write explicitly:
\begin{equation}\label{eq:p_non_nov}
P_{\text{non-nov}}(l)=\frac{ \int_0^{t_{\text{max}}} \left(1 + \frac{\rho t}{N_0} \right)^{\frac{2 \nu}{\rho} - 2} \left( 1-\left(1 + \frac{\rho t}{N_0} \right)^{\frac{\nu}{\rho} - 1}\right)^l  dt }{\int_0^{t_{\text{max}}} \left(1 + \frac{\rho t}{N_0} \right)^{\frac{\nu}{\rho} - 1} \left( 1-\left(1 + \frac{\rho t}{N_0} \right)^{\frac{\nu}{\rho} - 1}\right) dt}.
\end{equation}
\noindent The curves in Figs. \ref{fig:panel_non_nov_full},\ref{fig:panel_non_nov_wiki_and_gut}(a-b-e-f),\ref{fig:avalanches_umt_theo} and \ref{fig:scaling_umt} are obtained by performing a numerical integration of Eq.~(\ref{eq:pl}).  In the Supplementary Information (S7), we also provide an explicit series expansion for $P_{\text{non-nov}}(l)$.

The avalanche size distribution for novelties can be found in a completely analogous way, obtaining:
\begin{equation}
P_{\text{nov}}(l)=\frac{ \int_0^{t_{\text{max}}} p_{\text{new}}(t)^l ( 1-p_{\text{new}}(t))^2  dt }{\int_0^{t_{\text{max}}} p_{\text{new}}(t) ( 1-p_{\text{new}}(t)) dt},
\end{equation}
and, substituting the expression for $p_{\text{new}}(t)$:
\begin{equation}
P_{\text{nov}}(l)=\frac{ \int_0^{t_{\text{max}}} \left(1 + \frac{\rho t}{N_0} \right)^{\frac{l \nu}{\rho} - l} \left( 1-\left(1 + \frac{\rho t}{N_0} \right)^{\frac{\nu}{\rho} - 1}\right)^2  dt }{\int_0^{t_{\text{max}}} \left(1 + \frac{\rho t}{N_0} \right)^{\frac{\nu}{\rho} - 1} \left( 1-\left(1 + \frac{\rho t}{N_0} \right)^{\frac{\nu}{\rho} - 1}\right) dt}.
\end{equation}
We note that the avalanche size distributions depend on the ratios $\frac{N_0}{\rho}$ and $ \frac{\nu}{\rho} $, and on the sequence length $
t_{\text{max}}$. 
However, we can rescale the avalanche size $l$ to let curves obtained at different sequence lengths $t_{\text{max}}$ collapse into a unique curve, as shown in the next subsection.

In Fig.~\ref{fig:avalanches_umt_theo}, we contrast simulations and analytical results of the inter-times distribution of the UMT-E for different values of the ratio $\nu/\rho$ at fixed $N_0/\rho$,  (a), and for different values of $N_0/\rho$ at fixed ratio $\nu/\rho$ (b), 
at a fixed sequence length in both cases. 
We note that, as one can intuitively anticipate, the distribution features a support that enlarges when those ratios decrease. We report the corresponding analysis of the avalanches of novelties in the (S2) of the Supplementary Information.

\subsubsection{Avalanches statistics in the general UMT}

Building on the results of the previous section, to express the non-novelties avalanche size distribution, we consider $D$ as the independent variable in Eq.~(\ref{eq:pl}). We obtain $p_{\text{new}}(D)$ by substituting~\ref{eq:t_D_result} in~\ref{eq:heaps_generalUMT}, and we can thus write:
\begin{eqnarray}\label{eq:pl_generalUMT}
P_{\text{non-nov}}(l)&=&\frac{ \int_0^{D_{\text{max}}} p_{\text{new}}(D)^2 ( 1-p_{\text{new}}(D))^l  \frac{dt}{dD} dD}{\int_0^{D_{\text{max}}} p_{\text{new}}(D) ( 1-p_{\text{new}}(D)) \frac{dt}{dD} dD} = \nonumber \\
&&\nonumber \\
&=& \frac{ \int_0^{D_{\text{max}}} p_{\text{new}}(D)( 1-p_{\text{new}}(D))^l  d(D)}{\int_0^{D_{\text{max}}} ( 1-p_{\text{new}}(D)) dD}
\end{eqnarray}
Here $D_{\text{max}}=D(t_{\text{max}})$, and again we can perform a numerical integration.
We can proceed in an analogous way to obtain the expression for $P_{\text{nov}}(l)$. In the Supplementary Information (S4), we show that the above expression agrees with simulation results. In Supplementary Information (S5), we also compare the accuracy obtained using the approximate expression in Eq.~(\ref{eq:t_D_approxresult}) with that obtained using the asymptotic solution in~\cite{tria2014dynamics}. We show that both the expression in Eq.~(\ref{eq:t_D_approxresult}) and its counterpart obtained by imposing $D(1)=1$ achieve an accuracy comparable to that of the expression in Eq.~(\ref{eq:pl_generalUMT}). By contrast, the asymptotic solution in~\cite{tria2014dynamics} does not yield a reasonable approximation of the avalanche size statistics when $N_0$ exceeds a few units, as is the case in all the real-world systems we analyzed.

\subsection{Scaling relation in the model and in real data}

\begin{figure*}[htbp!]
	\centering
	\includegraphics[width=\linewidth]{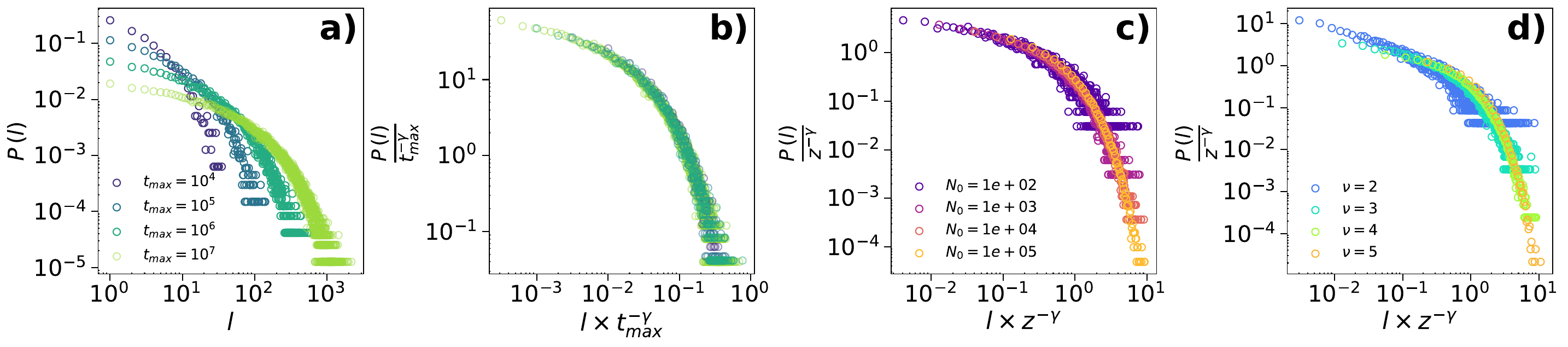}
	\caption{\textbf{Scaling laws in the avalanche size distribution for non-novelties in the UMT-E.} \textbf{a)} Avalanche size distribution of non-novelties for an instance of the same UMT-E process cut at different lengths. We set the parameters' values to $\nu=3,\rho=6,N_{0}=1000$. \textbf{b)} The same distributions in \textbf{a)}  rescaled so that they superimpose on a unique curve. \textbf{c)} The same distributions as in Fig.~\ref{fig:avalanches_umt_theo}({\bf b)}, ($\nu=3 $, $\rho=6$, and $N_{0}$ running across different orders of magnitude from $10^{2}$ to $10^{5}$) rescaled so that they superimpose on a unique curve. 
 \textbf{d)} The same distributions as in Fig.~\ref{fig:avalanches_umt_theo}({\bf a)}, ($N_{0}=1000 $, $\rho=6$, $\nu$ ranging in the integer values $[2,5]$),
 rescaled as in \textbf{c)}.}
	\label{fig:scaling_umt}
\end{figure*}

The results above show that the avalanche size distributions depend on the sequence length $t_{\text{max}}$.
We seek now a scaling relation to compare the avalanche size statistics at different sequence lengths. We observe that an interval between two successive novelties corresponds to an increase by one of the total number of distinct elements $D(t)$.
We can thus write, setting the identity $l \equiv \Delta t$:
\begin{equation}
\frac{1}{l} \simeq \frac{\Delta D(t)}{\Delta t} \simeq \frac{dD(t)}{dt}= \left(1+\frac{\rho}{N_0}t\right)^{\frac{\nu}{\rho}-1} 
\end{equation}
We thus make the ansatz:
\begin{equation}
\frac{1}{l_\text{max}}\sim \left(1+\frac{\rho}{N_0}t_\text{max}\right)^{\frac{\nu}{\rho}-1} \sim t_\text{max}^{\frac{\nu}{\rho}-1}
\end{equation}
In Fig.~\ref{fig:scaling_umt}(a) we show the inter-times distribution of sequences of different lengths from the UMT-E. In Fig.~\ref{fig:scaling_umt}(b) we show the same data as
in Fig.~\ref{fig:scaling_umt}(a) as a function of the rescaled variable $l \times t_{max}^{-\gamma}$, by defining $\gamma\equiv \frac{\nu}{\rho}-1$.
We can see that the curves at different sequence lengths perfectly collapse on a unique curve.
From the results in the above section, we can further observe that sequences generated by the UMT-E with the same Heaps' exponent, i.e., with the same ratio
$\frac{\nu}{\rho}$, and different $N_0$, can be rescaled on a unique curve by means of the variable $l \times z^{-\gamma}$, with $z\equiv \frac{\rho}{N_0} t_\text{max}$. In Fig.~\ref{fig:scaling_umt}(c), we show the same data as in Fig.~\ref{fig:avalanches_umt_theo}(b) as a function of the rescaled variable $l \times z^{-\gamma}$. We can see that the curves at different values of the ratio $\frac{N_0}{\rho} $ perfectly collapse on a unique curve. We further show in Fig.~\ref{fig:scaling_umt}(d) that the dependence on the ratio 
$\frac{\nu}{\rho}$ is not eliminated by rescaling the avalanche's length, but remains weak.

\begin{figure*}[htpb!]
\centering
\includegraphics[width=\linewidth]{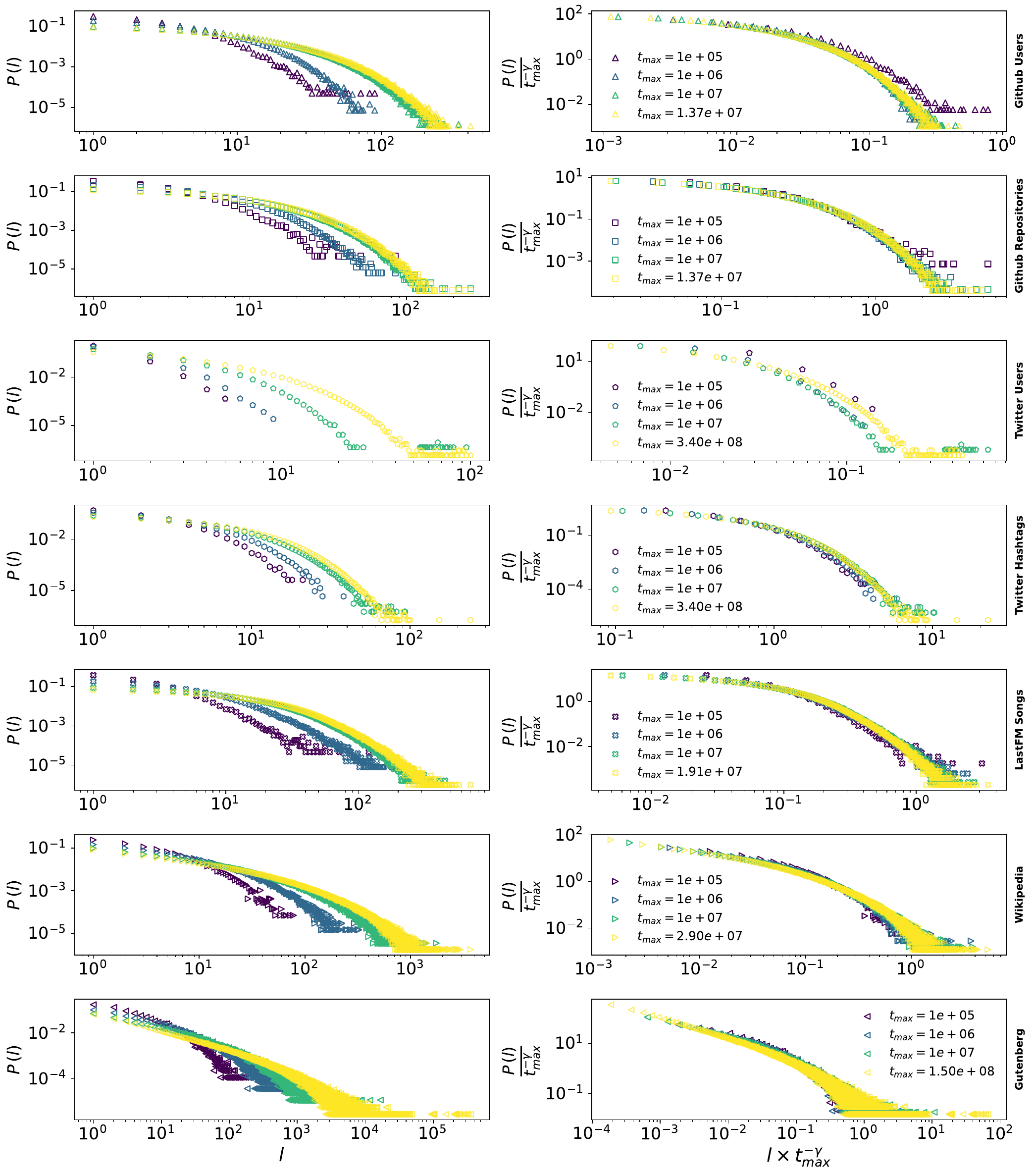}
\caption{\textbf{Scaling laws in the non-novelties avalanche size distributions for different sequences' lengths of each real-world dataset}. We consider the sequence of each dataset truncated at different lengths $t_{\max}$. Left: Avalanche size distributions
of non-novelties in each dataset for sequences of different lengths. Right: same distributions as on the right, where we adopted the scaling law derived for the UMT-E. The $\gamma$ exponent is fitted from the longest sequence in each dataset, as described in the main text.}
\label{fig:scaling_real_data}
\end{figure*}

In Fig. \ref{fig:scaling_real_data}, we report the same scaling applied to all the datasets, showing that all the innovation systems considered obey the scaling relation for different sequence lengths. For each dataset, we truncated the whole sequence at various lengths. We set $\frac{\nu}{\rho}$ to the value measured over the Heaps law of the entire sequence (we note that, for sufficiently long sequences, the fitted value $\frac{\nu}{\rho}$ does not depend on the sequence length).

\begin{figure}[h!]
	\centering
	\includegraphics[width =1\linewidth]{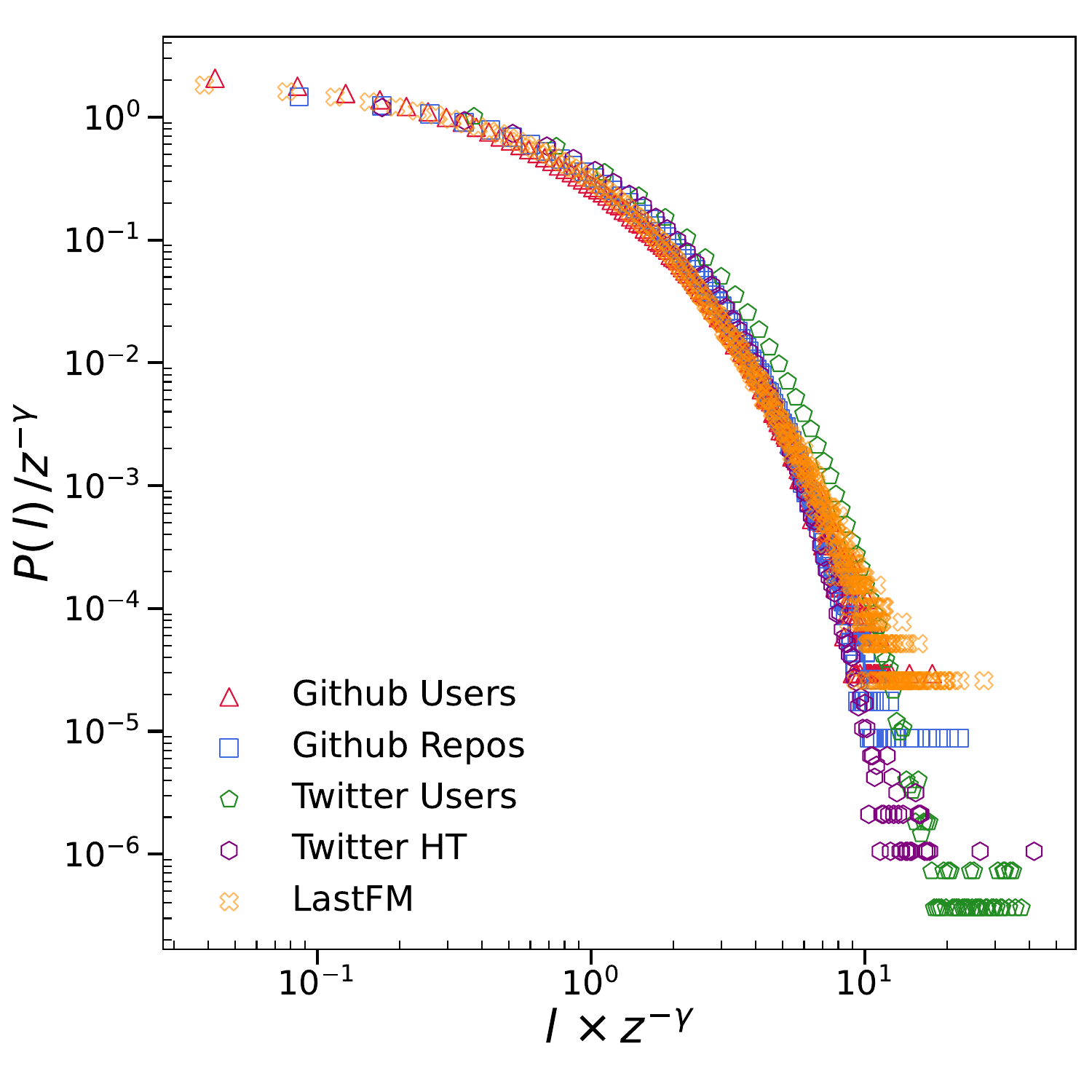}
	\caption{\textbf{Scaling laws in the non-novelties avalanche size distributions in the real-world datasets.} We here rescale the length $l$ of the non-novelties avalanches according to the variable  $z^{-\gamma}$. This rescaling allows the curves for different systems to almost collapse into one unique shape. We fit the $\gamma$ exponent and the ratio $N_0/\rho$ from the longest sequence in each dataset.}
	\label{fig:scaling_z}
\end{figure}

The seven considered datasets do not feature the same Heaps'  exponents. However, all the exponents of the five datasets whose avalanche size statistics  are well reproduced by a simple UMT model (LastFM, Github repositories, Twitter Hashtags, Github, and Twitter users) span a sufficiently small range, and the curves of the inter-time distributions 
approximately superimpose when plotted as a function of the variable $l \times z^{-\gamma}$ (Fig.~\ref{fig:scaling_z}).
  As for the ratio $\frac{\nu}{\rho}$, the ratio $\frac{N_0}{\rho}$ is estimated for each dataset by fitting the Heaps' law in Eq.~(\ref{eq:Dt_UMTE}), which gives the parameters reported in the captions of Figs.~\ref{fig:panel_nov_full} and~\ref{fig:panel_non_nov_full}.

\section*{Methods}
\subsection{Data}
Our analysis takes into account seven different data sets.

\noindent (i-ii) \textbf{Github} is a repository hosting service that allows many software developers to work on the same project. Users can register for this online platform and create new projects or interact with existing ones (provided they are public or have access to them). The interaction can range from creating a new branch of the code to creating or modifying the Wiki page related to the project, and so on. In our analysis, we do not distinguish between the different possible interactions within the repositories, and we focus instead on the time sequence of such interactions, so that the elements of the temporal sequence are the repositories themselves. The data we used contains the activity of GitHub users in January $2015$, with $1.4 \times 10^{7}$ actions performed over $1.5 \times 10^{6}$ repositories by $8.8 \times 10^{5}$ users. In this work, we have considered both the time series of the actions taken over the repositories and the time series of the users performing those actions. 

\noindent (ii-iv) \textbf{Twitter}, now referred to as  \textbf{X}, is a social networking service. It is one of the world's largest social media websites and the fourteenth-most visited website. Users can share short text messages, images, and videos in posts (formerly "tweets") and like or repost/retweet other users' content. X includes direct messaging, video and audio calling, bookmarks, lists and communities, and Spaces, a social audio feature. The dataset we used was collected daily in January 2013, recording $10\%$ of the users' activity. Tweets without "hashtags" have not been considered. The dataset thus consists of a sequence of time-ordered hashtags adopted by users, each identified by its own ID. Every hashtag is considered a separate element. We have considered the time series of hashtags within the same tweet in the order in which they appear. We restricted our analysis to the first week of data, and we considered both the time series of the hashtags and the time series of the users performing the corresponding tweets. In total, we have $3.47 \times10^{7}$ tweets containing $6.9 \times 10^{6}$ distinct hashtags shared by $1.28 \times 10^{7}$ different users.

\noindent (v) \textbf{LastFM} is a music website with a recommendation system. LastFM builds a detailed profile of each user's musical taste by recording details of the songs the user listens to, either from Internet radio stations, the user's computer, or many portable music devices. The data set we used contains the whole listening habits of 1000 users till May 5th, 2009, recorded in plain text form. It contains about $1.9 \times 10^{7}$ listened tracks with information on user, time stamp, artist, track-id, and track name, resulting in $1.08 \times 10^{6}$ different songs.

\noindent (vi) The \textbf{Gutenberg Project} is a Corpus of English literary texts derived from the Gutenberg Project e-book collection till February 2007, resulting in a set of about 4600 non-copyrighted e-books from which we have selected the 1000 longest, of various topics and including both prose and poetry. We preprocessed the Corpus in our analysis, adopting the following choices: we ignored capitalization. We took as distinct words sharing the same lexical root, i.e., we considered the word \textit{flower}  different from \textit{flowers}. We treated homonyms as the same word, for example, the verbal past perfect saw and the substantive saw. We removed punctuation marks.
The sequence obtained by stacking all 1000 books together comprises $1.58 \times 10^{8}$ words with a total vocabulary of $3.86 \times 10^{5}$ words. We stacked books in a random order, since information about each text's creation time is not available. 

\noindent (vii) The  \textbf{English Wikipedia} dataset we analyzed consists of the texts of 40000 different Wiki pages, randomly selected from the whole Wikipedia Corpus. We perform the same preprocessing as in the Gutenberg Project corpus.  The sequence is obtained by stacking all the pages in a unique text of $2.90 \times10^{7}$ words, where the number of distinct words is $6.73 \times 10^{5}$. 

\subsection{The mapping}
To test the hypothesis that the avalanche size statistics of the Gutenberg and the Wikipedia corpora reflect a \textit{convolution} between different semantic contexts, we reproduce this setting within our modeling scheme. We do that most simply, not aiming at giving a full modeling framework, but to show that accounting for the superposition of semantic contexts is crucial to understanding the dynamics
observed in those datasets. We then proceeded as follows:

\begin{enumerate}
\item For each Corpus, we fit the Heaps' law of each textual sequence, say $S_{C}^{i}$ with vocabulary size $D_{C}^{i}$ (each book for the Gutenberg corpus and each Wikipedia page for the Wikipedia corpus), with the analytical formula predicted for the UMT-E (Eq. \ref{eq:Dt_UMTE}) and estimate the ratios $\frac{\nu}{\rho} \equiv \alpha$ and $\frac{N_{0}}{\rho} \equiv \theta$.
From the parameters $\theta$ and $\alpha$, we obtained $\nu$ and $\rho$ as the two integers whose ratio $\frac{\nu}{\rho}$ is the closest to $\alpha$, in the ranges $\nu \in [1,99]$ and $\rho \in [\nu+1,100]$.
Once we fixed the value of $\rho$, we set $N_{0}$ as the closest integer to  $\theta \times \rho$.
\item We generate a sequence from UMT-E, say $S_{UMT-E}^{i}$, with the estimated parameters and of the same length of $S_{C}^{i}$, resulting in a vocabulary size $D_{UMT-E}^{i}$.
\item If $D_{UMT-E}^{i} > D_{R}^{i}$, we cut $S_{UMT-E}^{i}$ at the time $t$ such that $D_{UMT-E}^{i}(t+1)=D_{C}^{i}+1$, to have $D_{UMT-E}^{i}=D_{C}^{i}$.
\item We sort each word in $S_{C}^{i}$ and each element in $S_{UMT-E}^{i}$ according to their rank.
We replace each distinct element of $S_{UMT-E}^{i}$ with a word in $S_{C}^{i}$ according to their frequencies: we replace the most frequent element of $S_{UMT-E}^{i}$ with the most frequent word in $S_{C}^{i}$, the second most frequent element in $S_{UMT-E}^{i}$ with the second most frequent word in $S_{C}^{i}$ and so on.
\item Finally, for each Corpus, we stack together all the sequences $S_{UMT-E}^{i}$ in the same order as the corresponding sequences $S_{C}^{i}$.
\end{enumerate}

\subsection{The UMST model}
The UMST model introduced in \cite{tria2014dynamics} is a generalization of the UMT, where the concept of semantic groups is added. The dynamics is such to favor the stay within a semantic group,  though a fourth parameter $\eta$ belonging to the real interval $[0,1]$. The process works as follows: at $t=0$ the urn contains $N_{0}$ different elements divided in $\frac{N_{0}}{\nu+1}$ different semantic groups. We randomly extract the first element.
At each time step $t$ an element $s_{t}$ is drawn from the urn according to a probability proportional to its occurrencies inside the urn if it belongs to the same semantic group of $s_{t-1}$; it is instead drawn with a probability proportional to its occurrencies inside the urn times $\eta$ if it belongs to a semantic group different from $s_{t-1}$. The reinforcement and triggering rules are analogous to the ones adopted in the UMT model. If $s_{t}$ is a novelty, it is reinforced with $\rho$ copies inside the urn, and further enlarges the adjacent possible by introducing $\nu+1$ new distinct elements inside the urn. These $\nu+1$ new elements share a new semantic group. Only the reinforcement process occurs if $s_{t}$ is a non-novelty. In the $\nu < \rho $ case, in analogy with the UMT, if the last extracted element is a novelty, we can choose the reinforcement parameter to be $\tilde{\rho}$. Note, however, that for $\tilde{\rho}=\rho-(\nu+1)$ the UMST model is still non exchangeable.

\section*{Conclusions}

Over the past years, several studies have advanced the identification of significant statistical signatures of innovation processes~\cite{cattuto2007heaps,tria2014dynamics,gerlach2014}. 
This work introduces the avalanche statistics as key observables for characterizing their dynamics, complementing the well-established Zipf's, Heaps', and Taylor's laws~\cite{heaps1978information,zipf35,taylor1961aggregation}.

Within the framework of the urn model with triggering (UMT) ~\cite{tria2014dynamics}, we derive exact expressions for the avalanche size distributions of novelties and non-novelties, which reproduce empirical observations across diverse innovation-related datasets, including collaborative platforms, social media, cultural production, and natural language corpora.
At the same time, we extend the existing characterization of Heaps' law within the UMT framework~\cite{tria2014dynamics,tria2018zipf}, emphasizing the role of the initial size of the adjacent possible~\cite{kauffman1996investigations}. 

We establish an equivalence between the UMT, its exchangeable version (UMT-E), and the urn model with semantic triggering (UMST)~\cite{tria2014dynamics}, i.e.,  when tuned to generate the same Heaps law, these models produce identical avalanche statistics.
We recall that the UMST model was devised to reproduce correlations in the occurrence of related items, as quantified, for instance, by entropic measures~\cite{tria2014dynamics}. These correlations provide further evidence for the adjacent possible expansion mechanism at work and are not accounted for by the simple UMT model. In the Supplementary Information (S6), we discuss a signature of non-exchangeability provided by the Heaps law.

The equivalence of the three models with respect to the avalanche size statistics sets a direct connection between Heaps' law and avalanche dynamics. 
However, the avalanche size statistics, especially for non-novelties, contain more information than Heaps' law alone. Empirical analyses also reveal deviations from model predictions, suggesting that avalanche size statistics can distinguish between dynamics that can be effectively described as a single collective process (and thus obey the UMT predictions) and those that emerge from the superposition of individual contributions,  uncovering dynamical features that go beyond those described by the Heaps law.
Indeed, both the Wikipedia and Gutenberg corpora exhibit heavy-tailed interevent time distributions, consistent with the aggregation of individual contributions. However, only Gutenberg shows a clear power-law regime for small avalanche sizes, reflecting the higher internal complexity of books, where multiple topics and dynamics coexist. The similarity between the avalanche statistics of Wikipedia and those of individual books within the Gutenberg corpus supports this interpretation.

From the UMT-E analytical results, we provide a scaling relation that allows us to superimpose avalanche distributions from the same system but corresponding to different sequence lengths. All the datasets considered obey this, revealing a form of universality in the dynamics of innovation. Moreover, we derive a scaling relation that extends the collapse on a unique curve to nearly all considered datasets, except for the two natural language corpora, which deviate from the model predictions. This result highlights the roles of the initial repertoire size and the rate of the adjacent possible expansion for each system.

Overall,  the avalanche framework provides insights that deepen our understanding of the mechanisms driving innovation. We believe it offers a foundation for future theoretical and empirical studies across innovation systems.

\subsection{Data Availability}
The data used in this manuscript are available at \url{https://figshare.com/articles/dataset/datasets_avalanches/30186733} and have been obtained and used according to their terms and conditions.
\subsection{Code Availability}
All the codes used to process and analyze the data and numerically realize the models can be found at \cite{F.Santoro}.

\clearpage
\onecolumngrid
\section{Supplementary Information for "Universalities in the Avalanche Dynamics of Novelties and Non-Novelties and some Notes on the Heaps law"}
\section{S1. Heaps' law Fit for Real Datasets}
\begin{figure}[htbp]
        \includegraphics[width=\linewidth]{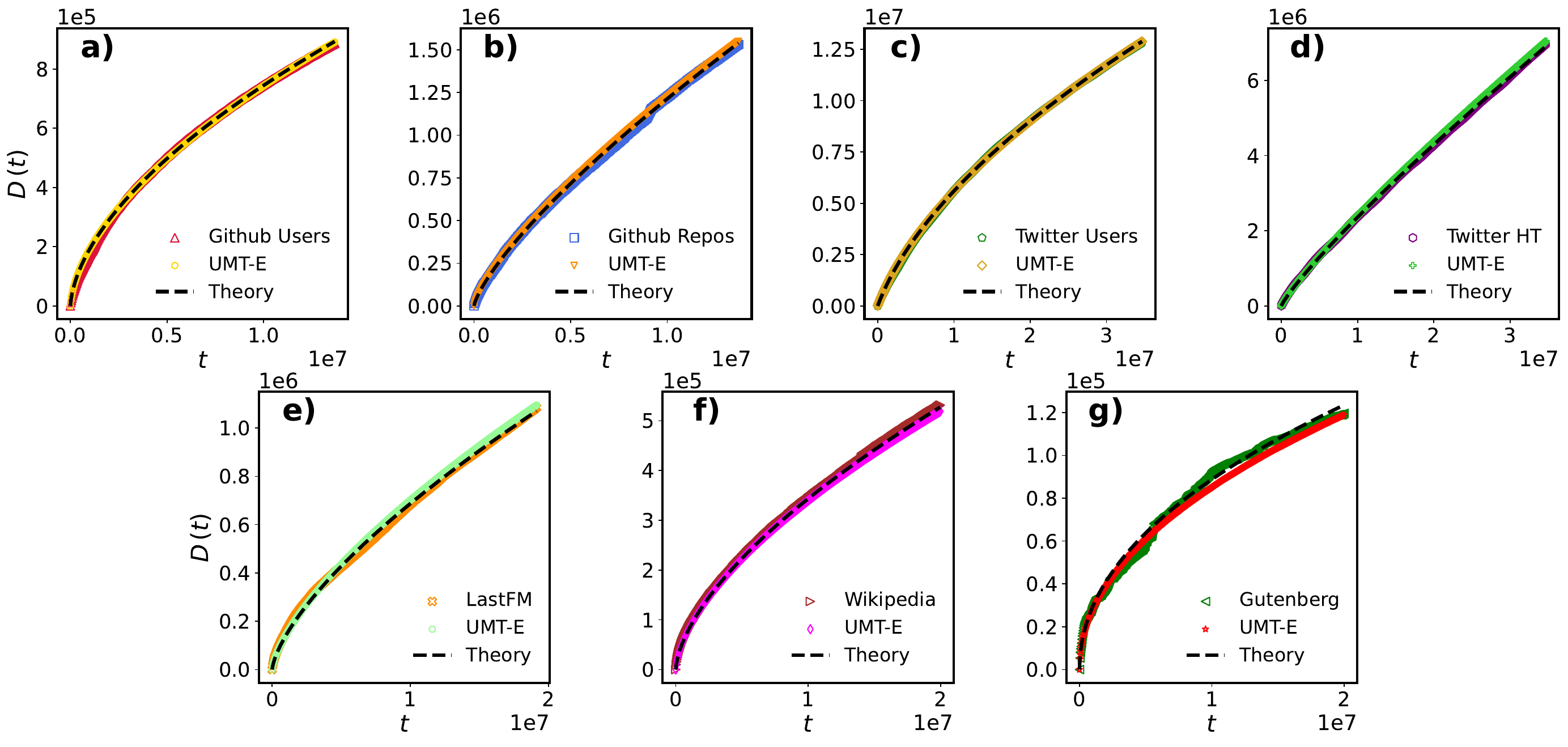}
        \caption{\textbf{The Heaps law in real datasets.} Comparison between the Heaps Law of real datasets and the analytical UMT-E Heaps law that best fits each dataset, along with the corresponding numerical realization of the process.}
        \label{fig:heaps_datasets}
\end{figure}

We predict the avalanche size distribution of novelties and non-novelties in each real-world system from the UMT-E with the parameters $(\nu,\rho,N_{0})$ that best fit the Heaps law of each considered system. With these parameters, we derived the theoretical analytical expression for the avalanche distributions, which we call 'theory' in all the figures, and produced numerical realizations of the corresponding process from which we computed the empirical avalanche distributions (what we call 'UMT-E' in all the figures).
For each dataset, we have fitted its Heaps law with the formula:
\begin{eqnarray}
D(t) &= \nonumber\frac{\theta}{\alpha} \left(1 + \frac{ t}{\theta} \right)^{\alpha} - \frac{\theta}{\alpha} 
\end{eqnarray}
where $\theta=\frac{N_{0}}{\rho}$ and $\alpha=\frac{\nu}{\rho}$. We have then chosen $\nu$ and $\rho$ as the integers whose ratio is the closest to $\alpha$. This procedure can lead to an arbitrary precision, and we have chosen $\nu \in [1,99]$ and $\rho \in [\nu+1,100]$. Once $\nu$ and $\rho$ are obtained, we have estimated $N_{0}$ as the integer closest to the quantity $\theta\times\rho$. In Fig. \ref{fig:heaps_datasets} we show the Heaps law of all the datasets, the theoretical UMT-E Heaps law, and the Heaps law of the realization produced according to the estimated parameters.
For the sake of completeness, we report all the parameters in Table \ref{tab:dataset_params}.
\begin{table}[htbb!]
\centering
\caption{Datasets parameters}
\label{tab:dataset_params}
\begin{tabular}{|l|c|c|c|c|}
\hline
\textbf{Dataset} & \textbf{$\nu$} & \textbf{$\rho$} & \textbf{$N_0$} &
\textbf{$t_{max}$}\\
\hline
Github Users & 4 & 7 & 45,468 & $1.37 \times 10^{7}$\\
\hline
Github Repositories & 22 & 29 & 14,640 & $1.37 \times 10^{7}$ \\
\hline
Twitter Users & 16 & 29 & 42,183,043 & $3.4 \times 10^{8}$ \\
\hline
Twitter Hashtags & 19 & 22 & 1,995 & $3.4 \times 10^{8}$ \\
\hline
LastFM & 28 & 41 & 27,200 & $1.91 \times 10^{8}$\\
\hline
Wikipedia Corpus\textsuperscript{*} & 31 & 50 & 25,000 & $2.91 \times 10^{7}$ \\
\hline
Gutenberg Corpus\textsuperscript{*} & 9 & 19 & 5,890 & $1.58 \times 10^{8}$ \\
\hline
\end{tabular}
\begin{tablenotes}
\small
\item[*] *For computational reasons, UMT-E sequences were limited to 2.0 $\times 10^{7}$ elements for these parameter sets.
\end{tablenotes}
\end{table}

\section{S2. Avalanches of Novelties distributions for the UMT-E}
\begin{figure}[htbp]
        \includegraphics[width=\linewidth]{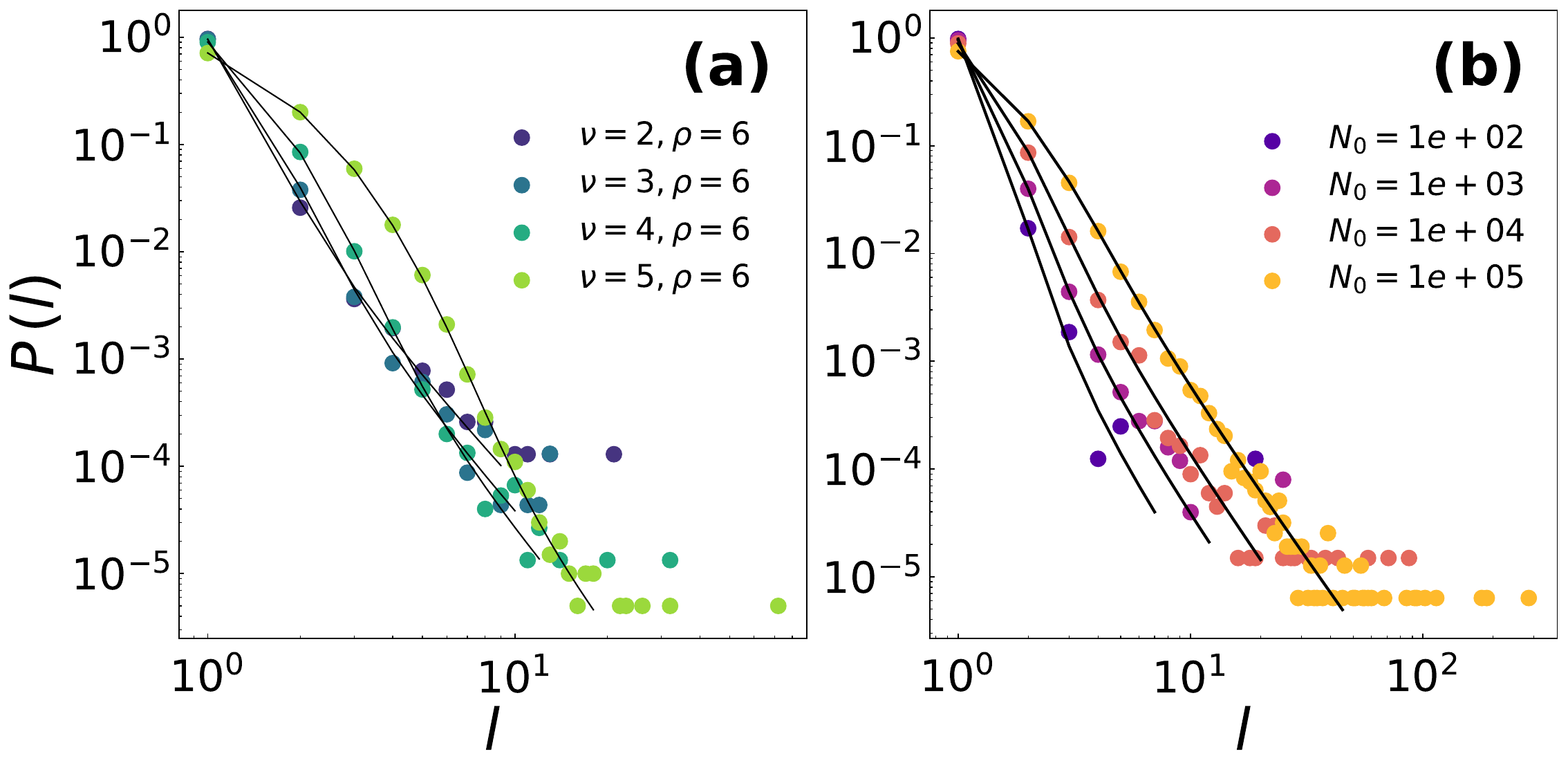}
        \caption{\textbf{Avalanche size distributions for Novelties in the UMT-E.} \textbf{a)} Realizations of length $10^{6}$ along with the theoretical predictions with $\rho=6$ and $\nu$ in the integer range $[2,5]$. $N_{0}=10^{3}$ for all the sequences.  \textbf{b)} Realizations of length $10^{6}$ along with the theoretical predictions for $(\nu=3,\rho=6)$. $N_{0}$ spans across the orders of magnitudes in the range $[10^{2},10^{5}]$.}
        \label{fig:avalanches_umt_e}
\end{figure}

In Fig. \ref{fig:avalanches_umt_e} we show the analysis on the avalanches of novelties of the UMT-E model.  By fixing $(\rho=6,N_{0}=1000)$ we see that increasing $\nu$ widens the support of the distribution, since larger avalanches become more probable to occur. In this sense, $\nu$ has the same role as $\rho$ in the avalanches of non-novelty distributions: increasing $\nu$ increases the probability of extracting a novel element, allowing also larger avalanches of novelties to form. Increasing $N_{0}$ also widens the support of the distribution. However, we remark that even though both increasing $N_{0}$ and $\nu$ favor innovations, their tuning reflects two different mechanisms: while increasing $\nu$ directly expands the adjacent possible,  increasing $N_{0}$ sets the initial condition of the space of possibilities.
\clearpage

\section{S3. Written Texts and Wikipedia Pages Analysis}
In this section, we present supplementary analyses performed on individual written texts and Wikipedia pages to compare and understand how the analytical results obtained for the UMT-E model reproduce the Heaps law and the avalanches of novelties and non-novelties distributions observed in real data.
\subsection{Written Texts Analysis}
\begin{figure}[htbp]
        \includegraphics[width=\linewidth]{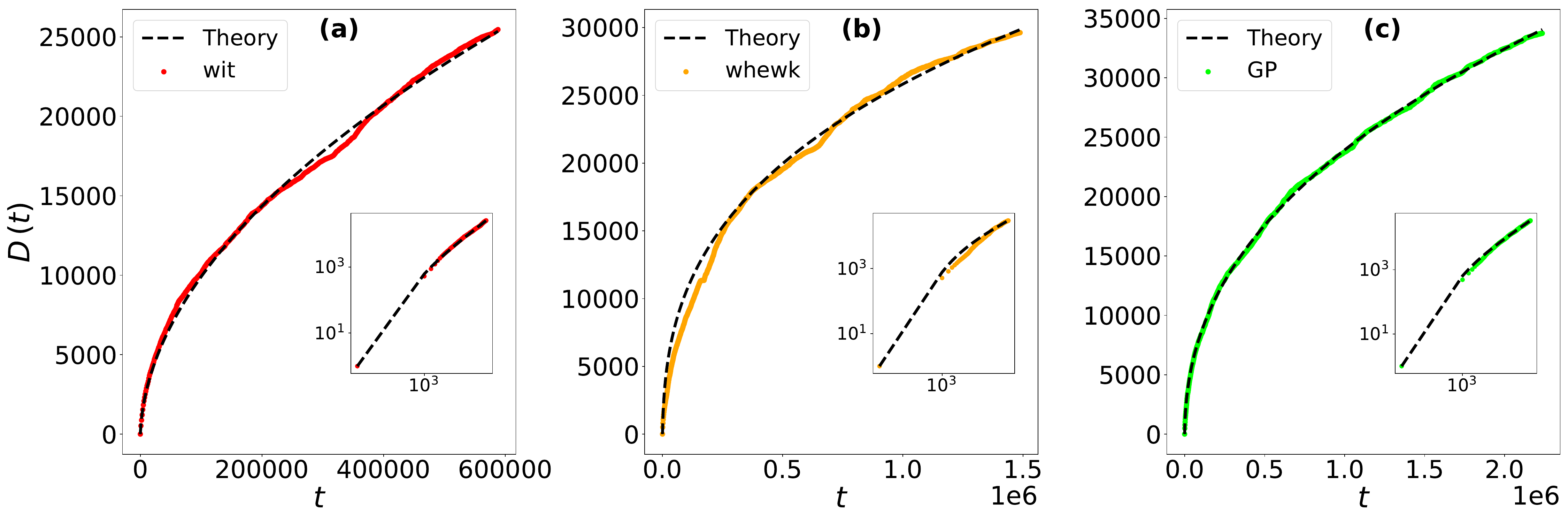}
        \caption{\textbf{Heaps Law of the UMT-E that best fits each written text.} \textbf{a)} Narrative and Legendary Poems by John Greenleaf Whittier (red plot).  \textbf{b)} Literary Friends and Acquaintances by William Dean Howells (orange plot). \textbf{c)} The Translation of a Savage of Gilbert Parker (lime plot).}
        \label{fig:heaps_books}
\end{figure}

\begin{figure}[htbp]
        \includegraphics[width=\linewidth]{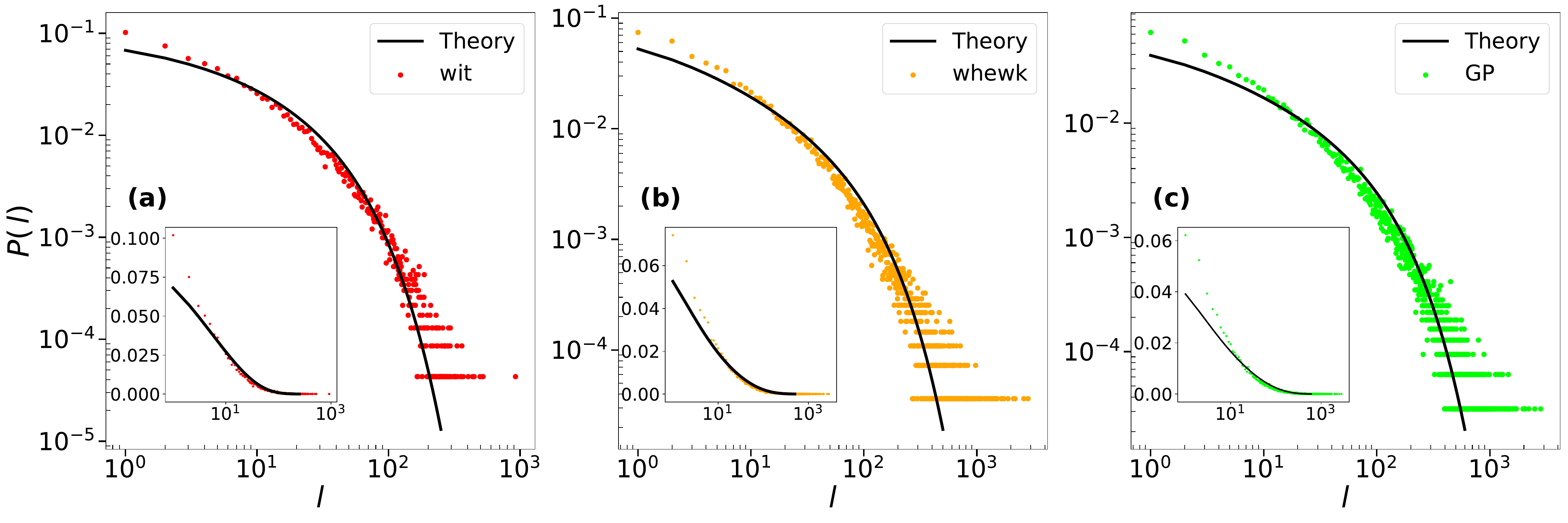}
        \caption{\textbf{Avalanche size distributions for non-novelties in Written Texts. } Theoretical curves (continuous black lines) are obtained by fitting the UMT-E model Heaps law to the Heaps law of each written text. Scattered colored plots represent the avalanche size distributions of non-novelties for each considered written text. \textbf{a)} Narrative and Legendary poems by John Greenleaf Whittier (red scattered plot).  \textbf{b)} Literary Friends and Acquaintances by William Dean Howells (orange scattered plot). \textbf{c)} The Translation of a Savage of Gilbert Parker (lime scattered plot).}
        \label{fig:delays_books}
\end{figure}

\begin{figure}[htbp]
        \includegraphics[width=\linewidth]{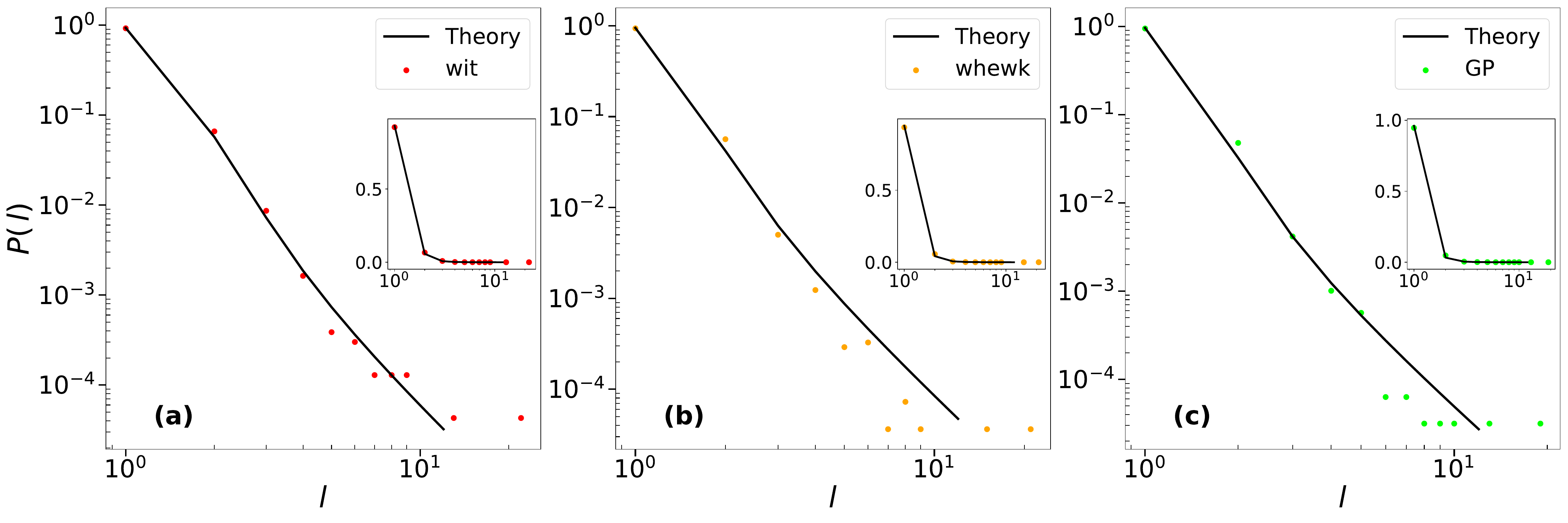}
        \caption{\textbf{Avalanches size distributions for novelties in Written Texts}. Theoretical curves (continuous black lines) are obtained by fitting the UMT-E model Heaps law to the Heaps law of each written text. Scattered colored plots represent the avalanche size distributions of Novelties for each considered written text. \textbf{a)} Narrative and Legendary Poems by John Greenleaf Whittier (red scattered plot).  \textbf{b)} Literary Friends and Acquaintances by William Dean Howells (orange scattered plot). \textbf{c)} The Translation of a Savage of Gilbert Parker (lime scattered plot).}
        \label{fig:avalanches_books}
\end{figure}

Here we report the analysis made on three written texts, randomly chosen among the $100$ longest in the Gutenberg corpus,  and their comparison with the UMT-E predictions. In Fig. \ref{fig:heaps_books}, we compare the experimental Heaps and the UMT-E Heaps that best fit it. In Fig. \ref{fig:delays_books} we show the comparison between the experimental intermission times distributions and their corresponding UMT-E predictions with the values of $(\nu,\rho,N_{0})$ derived from the fits of the Heaps law. In all the cases we have analyzed, the theoretical distribution predicts lower probability values for small inter-event times and slightly less fat tails than those observed in the written texts.
In Fig. \ref{fig:avalanches_books} we show the comparison between the predictions for the avalanche size distributions for novelties and the experimental observations with the same values of $(\nu,\rho,N_{0})$ mentioned above: in all the considered cases we find a good agreement between theoretical predictions and the experimental observations. 
In Table \ref{tab:literary_works} we resume the estimated parameters.
\begin{table}[htbp]
\centering
\caption{Parameters for literary works}
\label{tab:literary_works}
\begin{tabular}{|l|l|c|c|c|c|}
\hline
& \textbf{Work} & \textbf{$\nu$} & \textbf{$\rho$} & \textbf{$N_0$} & \textbf{Length} \\
\hline
\textbf{a)} & Narrative and Legendary poems by John Greenleaf Whittier & 50 & 97 & 22,911 & 588765 \\
\hline
\textbf{b)} & \begin{tabular}[c]{@{}l@{}}Literary Friends and Acquaintance\\by William Dean Howells\end{tabular} & 30 & 85 & 50,553 & 1485384 \\
\hline
\textbf{c)} & \begin{tabular}[c]{@{}l@{}}The Translation of a Savage \\by Gilbert Parker\end{tabular} & 35 & 82 & 29,258 & 2234347 \\
\hline
\end{tabular}
\end{table}

\begin{figure}[htbp]
        \includegraphics[width=\linewidth]{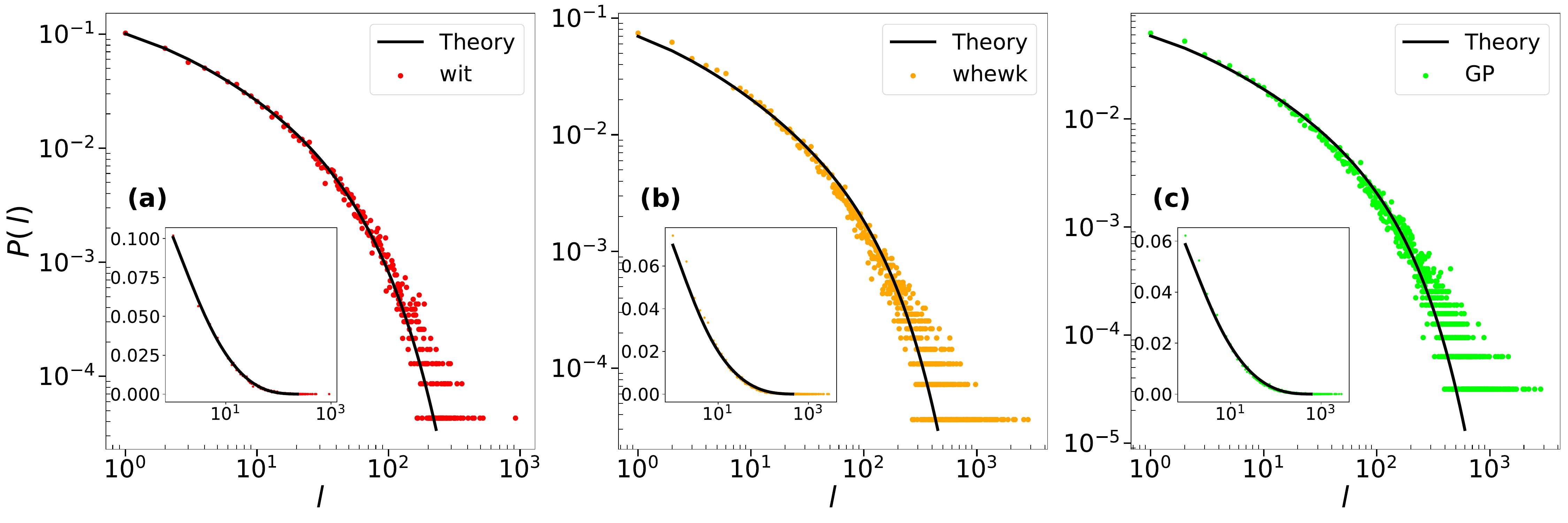}
        \caption{\textbf{Avalanche size distributions for non-novelties in Written Texts and corresponding theoretical predictions of the UMT-E model fitted on them.} Theoretical curves (continuous black lines) are obtained by fitting the theoretical UMT-E model avalanches of non-novelties size distribution directly on the experimental avalanches size distributions for non-novelties (scattered colored plots). \textbf{a)} Narrative and Legendary poems by John Greenleaf Whittier (red scattered plot).  \textbf{b)} Literary Friends and Acquaintances by William Dean Howells (orange scattered plot). \textbf{c)} The Translation of a Savage of Gilbert Parker (lime scattered plot).}
        \label{fig:delays_books_forced}
\end{figure}

In Fig. \ref{fig:delays_books_forced} we show that it is always possible to find a triple $(\nu,\rho,N_{0})$, different from the one which fits the Heaps law, that satisfactorily reproduces the intermission times distribution. In Table \ref{tab:literary_works_forced} we resume these parameters. 

\begin{table}[htbp]
\centering
\caption{Parameters for literary works fitted on the delays}
\label{tab:literary_works_forced}
\begin{tabular}{|l|l|c|c|c|c|}
\hline
& \textbf{Work} & \textbf{$\nu$} & \textbf{$\rho$} & \textbf{$N_0$} & \textbf{Length} \\
\hline
\textbf{a)} & Narrative and Legendary poems by John Greenleaf Whittier & 23 & 73 & 127020 & 588765 \\
\hline
\textbf{b)} & \begin{tabular}[c]{@{}l@{}}Literary Friends and Acquaintance\\by William Dean Howells\end{tabular} & 33 & 100 & 135000 & 1485384 \\
\hline
\textbf{c)} & \begin{tabular}[c]{@{}l@{}}The Translation of a Savage \\by Gilbert Parker\end{tabular} & 29 & 98 & 151900 & 2234347 \\
\hline
\end{tabular}
\end{table}
\clearpage
\subsection{Wikipedia Pages}
\begin{figure}[htbp]
        \includegraphics[width=\linewidth]{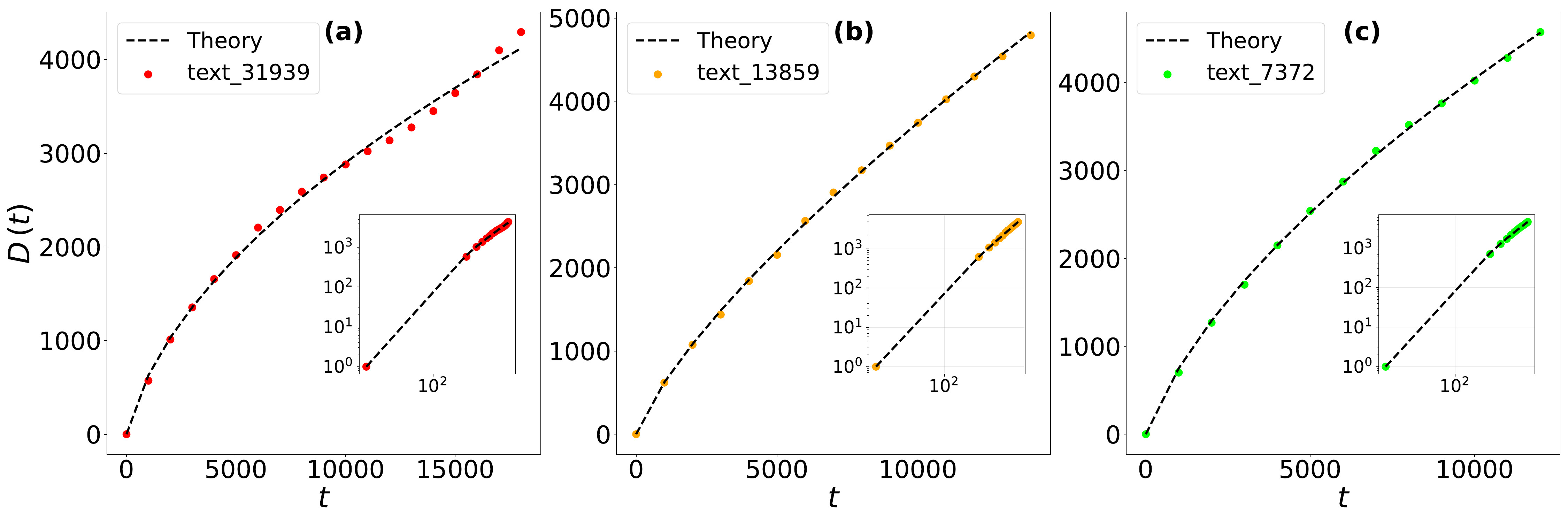}
        \caption{\textbf{Heaps law of the UMT-E that best fits each Wikipedia page.} \textbf{a)} text 31939 (red scattered plot). \textbf{b)} text 13859 (orange scattered plot). \textbf{c)} text 7372 (lime scattered plot). }
        \label{fig:heaps_wiki_pages}
\end{figure}

\begin{figure}[htbp]
        \includegraphics[width=\linewidth]{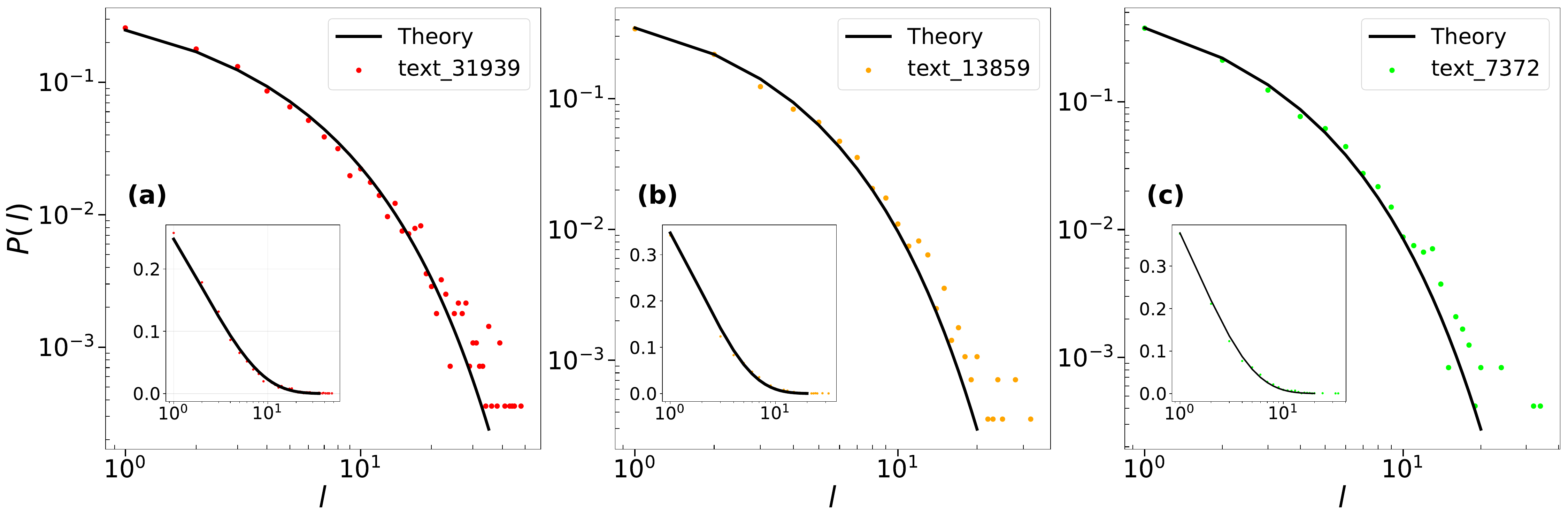}
        \caption{\textbf{Avalanche size distributions for non-novelties in Wikipedia Pages.} Theoretical curves (continuous black lines) are obtained by fitting the UMT-E model Heaps law to the Heaps law of each written text. Scattered colored plots represent the avalanche size distributions of non-novelties for each considered Wikipedia page. \textbf{a)} text 31939 (red scattered plot). \textbf{b)} text 13859 (orange scattered plot). \textbf{c)} text 7372 (lime scattered plot).}
        \label{fig:delays_wiki}
\end{figure}

\begin{figure}[htbp]
        \includegraphics[width=\linewidth]{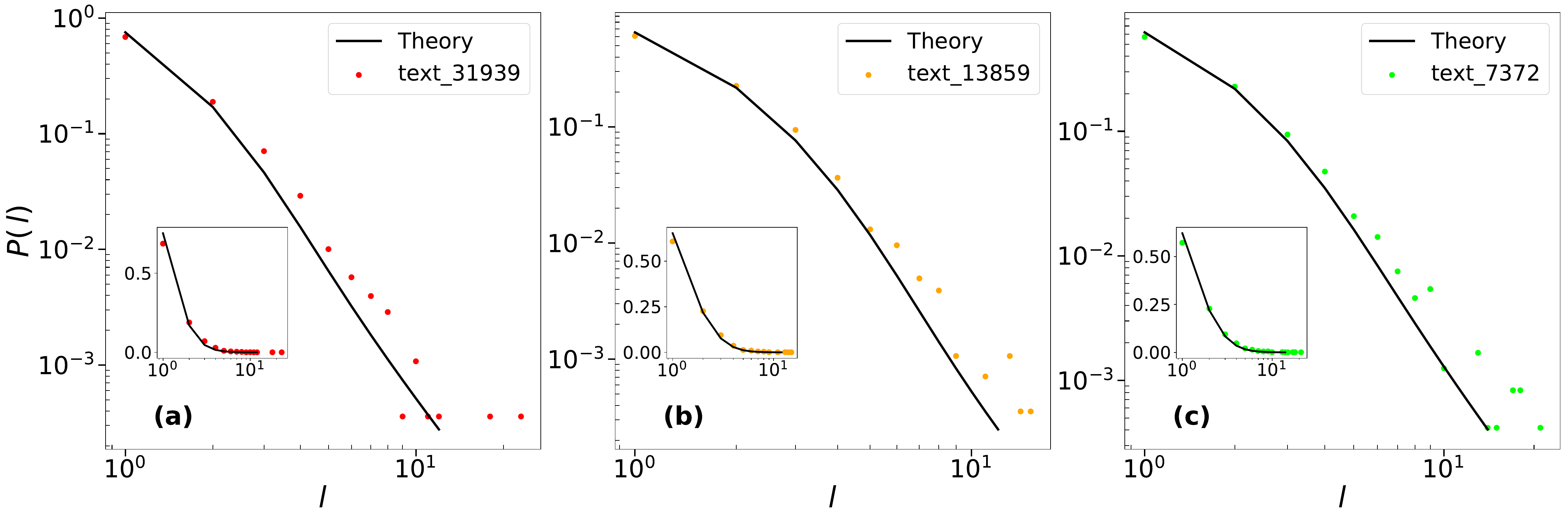}
        \caption{\textbf{Avalanche size distributions for novelties in Wikipedia Pages}. Theoretical curves (continuous black lines) are obtained by fitting the UMT-E model Heaps law to the Heaps law of each written text. Scattered colored plots represent the avalanche size distributions of novelties for each considered Wikipedia page. \textbf{a)} text 31939 (red scattered plot). \textbf{b)} text 13859 (orange scattered plot). \textbf{c)} text 7372 (lime scattered plot). }
        \label{fig:avalanches_wiki_pages}
\end{figure}

In this section, we follow the same methodology adopted for the written texts: we have randomly considered three pages among the 100 longest Wikipedia pages, and for each page, we compare the experimental Heaps law with the UMT-E Heaps law that best fits it in \ref{fig:heaps_wiki_pages}. 
In Fig. \ref{fig:delays_wiki} we show the comparison between the experimental intermission times distributions and their corresponding UMT-E predictions with the values of $(\nu,\rho,N_{0})$ derived from the fits of the Heaps law. In all the cases we have analyzed, the theoretical predictions from UMT-E agree with the observed empirical behavior. In Fig. \ref{fig:avalanches_wiki_pages} we show the corresponding avalanche size distributions for novelties. Table \ref{tab:wiki_pages} resumes the estimated parameters.
\begin{table}[htbp]
\centering
\caption{Parameters for Wikipedia Pages}
\label{tab:wiki_pages}
\begin{tabular}{|l|l|c|c|c|c|}
\hline
& \textbf{Page} & \textbf{$\nu$} & \textbf{$\rho$} & \textbf{$N_0$} & \textbf{Length} \\
\hline

\textbf{a)} & text 31939 & 27 & 50 & 11,716 & 18150 \\
\hline
\textbf{b)} & \begin{tabular}[c]{@{}l@{}} text 13859 \end{tabular} & 56 & 75 & 5283 & 14160 \\
\hline
\textbf{c)} & \begin{tabular}[c]{@{}l@{}} text 7372 \end{tabular} & 51 & 86 & 36062 & 12068 \\
\hline
\end{tabular}
\end{table}

\section{S4. Calculation of the Heaps law in the UMT model and the intermission times distribution}
\begin{figure}[htbp]
        \includegraphics[width=0.9\linewidth]{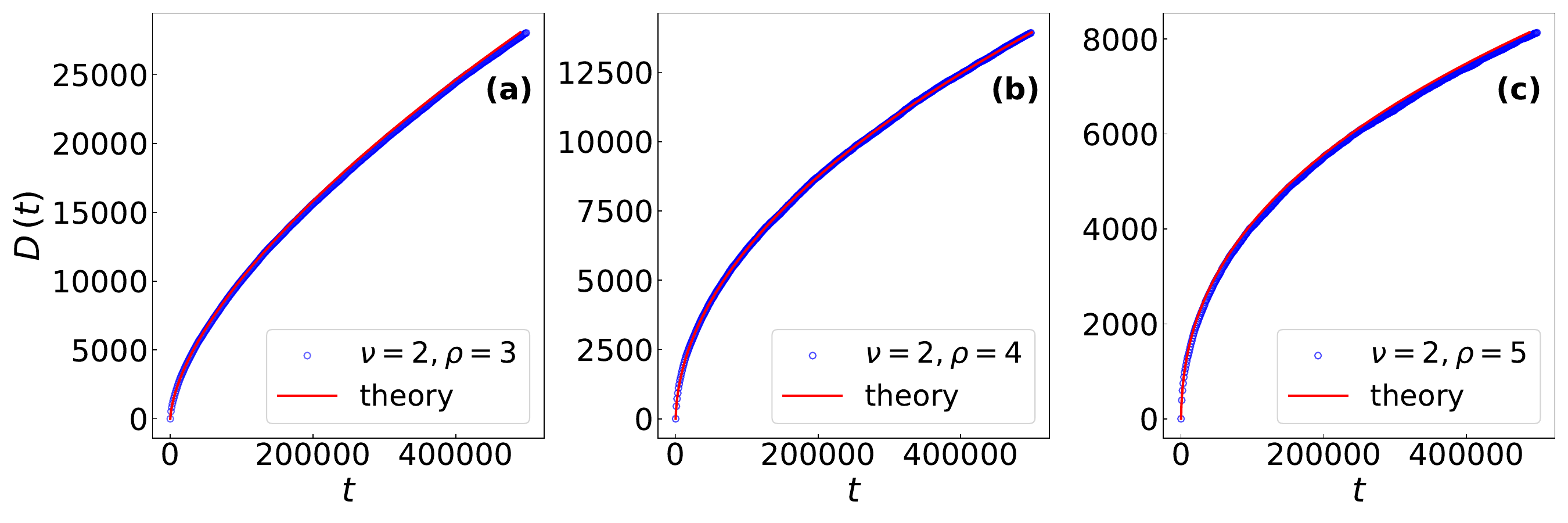}
        \caption{\textbf{Heaps Law for the UMT model.} Comparison between the theoretical predictions and corresponding realizations. \textbf{a)} $(\nu=2,\rho=3,N_{0}=1000)$. \textbf{b)} $(\nu=2,\rho=4,N_{0}=1000)$. \textbf{c)} $(\nu=2,\rho=5,N_{0}=1000)$. All the realizations have length $5\times10^{5}$. }
        \label{fig:heaps_theo_umt}
\end{figure}

We consider the general UMT model with $\Tilde{\rho}\geq 0$. The continuos time equation for  $D(t)$ reads:
\begin{equation}
\frac{dD(t)}{dt} = \frac{N_0 + \nu D(t)}{N_0 + \rho t + a D(t)}
\end{equation}
with $a \equiv \Tilde{\rho} + \nu+1 -\rho$, and for $a=0$ we recover the UMT-E, while for $a=\nu+1$ we recover the original UMT model.
Instead of solving this equation, we solve 
\begin{equation}
\frac{dt(D)}{dD} = \frac{N_0 + \rho t (D)+ a D}{N_0 + \nu D}
\end{equation}
that we can write as
\begin{equation}
\frac{dt(D)}{dD} - \frac{\rho }{N_0 + \nu D} t (D)=  \frac{N_0 + a D}{N_0 + \nu D}
\end{equation}
putting in evidence its nature as a non-homogeneous first-order linear differential equation.
Solving by standard methods, we write the solution for the homogeneous equation:
\begin{equation}
{t}_{\text{om}} = C\left( \frac{N_0 + \nu D}{N_0 }\right)^{\frac{\rho}{\nu}}
\end{equation}
and a particular solution for the non-homogeneous:
\begin{equation}
{t}_{\text{p}} = 	- a\left(\frac{N_{0}+\nu D}{\nu(\rho-\nu)} \right) -\frac{N_{0}}{\rho} \left( \frac{\nu-a}{\nu} \right)
\end{equation}
\noindent By finally imposing the initial condition $t(0)=0$, we obtain the solution
\begin{eqnarray}\label{eq:SI_tD}
t(D)&= &\frac{N_{0}}{\rho} \left(1 + \frac{a}{\rho-\nu} \right) \left(\frac{N_{0}+\nu D}{N_{0}} \right)^{\frac{\rho}{\nu}}  + \\
&- &a\left(\frac{N_{0}+\nu D}{\nu(\rho-\nu)} \right) -\frac{N_{0}}{\rho} \left( \frac{\nu-a}{\nu} \right) \nonumber
\end{eqnarray}
\noindent
The solution in this form is suitable for predicting the avalanche size distributions of novelties and non-novelties. For instance, for the avalanches of non-novelty distributions, it is only needed the numerical integration of 
\begin{eqnarray}\label{eq:pl_generalUMT}
P_{\text{non-nov}}(l)&=&\frac{ \int_0^{D_{\text{max}}} p_{\text{new}}(D)^2 ( 1-p_{\text{new}}(D))^l  \frac{dt}{dD} dD}{\int_0^{D_{\text{max}}} p_{\text{new}}(D) ( 1-p_{\text{new}}(D)) \frac{dt}{dD} dD} = \nonumber \\
&&\nonumber \\
&=& \frac{ \int_0^{D_{\text{max}}} p_{\text{new}}(D)( 1-p_{\text{new}}(D))^l  d(D)}{\int_0^{D_{\text{max}}} ( 1-p_{\text{new}}(D)) dD}
\end{eqnarray}
We show the perfect agreement between the solution and corresponding realizations of the process in Fig. \ref{fig:heaps_theo_umt}.

\begin{figure}[htbp]
        \includegraphics[width=0.9\linewidth]{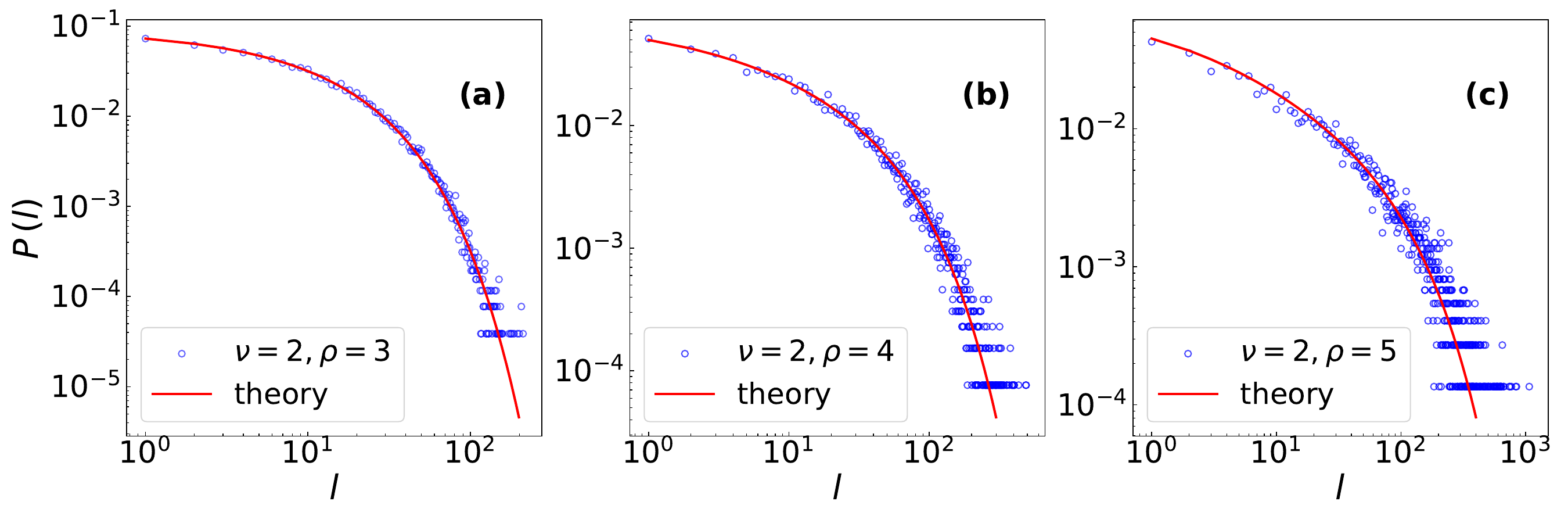}
        \caption{\textbf{Avalanches size distributions for non-novelties for the UMT model.} Comparison between the theoretical predictions and corresponding realizations. \textbf{a)} $(\nu=2,\rho=3,N_{0}=1000)$. \textbf{b)} $(\nu=2,\rho=4,N_{0}=1000)$. \textbf{c)} $(\nu=2,\rho=5,N_{0}=1000)$. All the realizations have length $5 \times 10^{5}$. }
        \label{fig:delays_UMT}
\end{figure}

In Fig. \ref{fig:delays_UMT} we consider the same relizations of the process and show the perfect agreement with the analytical avalanches of non-novelties distribution for the UMT model.  

\section{S5. Approximate solutions for Heaps' law in the general UMT model and the relative intermission times distribution}
\begin{figure}[htbp]
        \includegraphics[width=\linewidth]{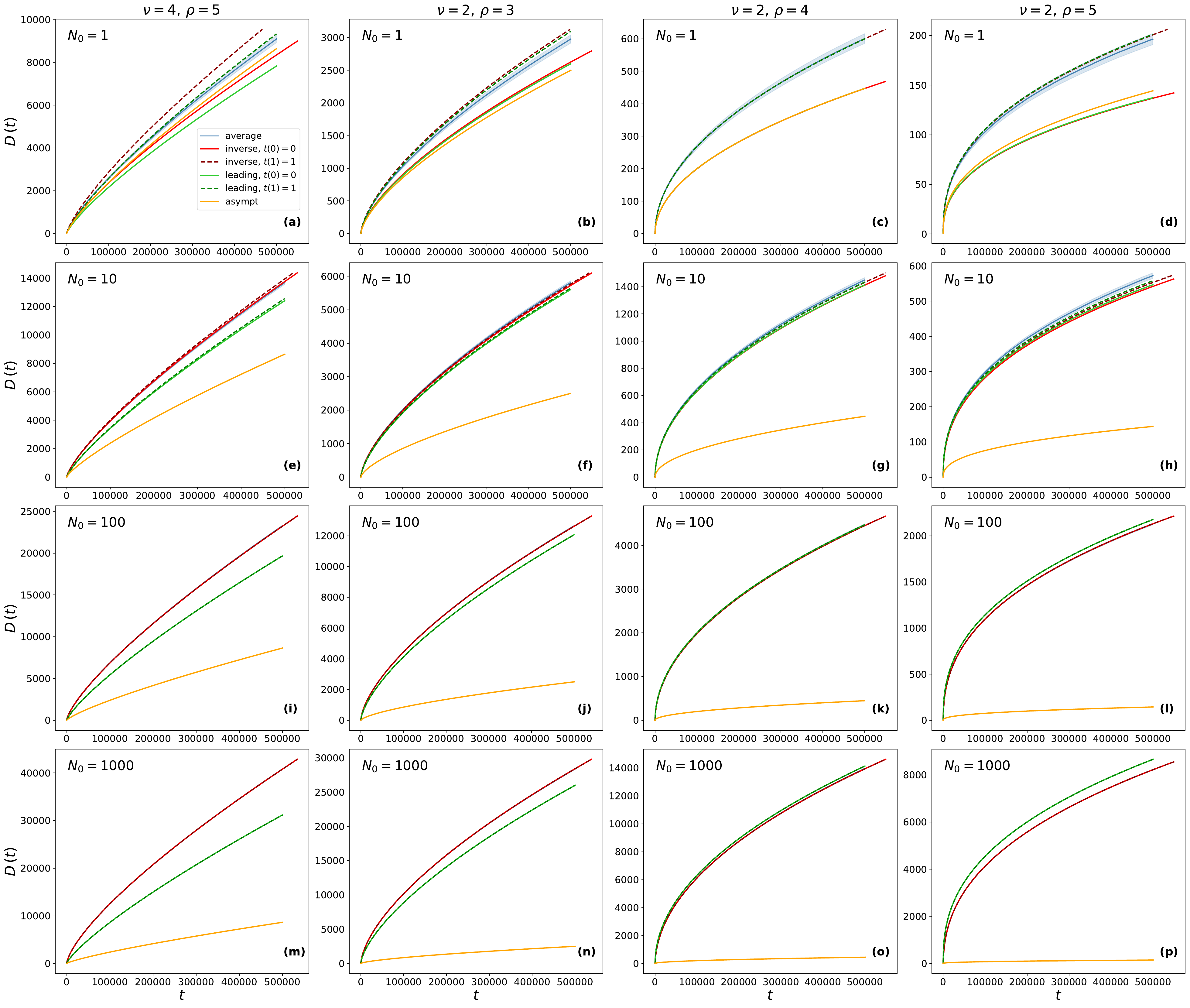}
        \caption{\textbf{Heaps' Law in the UMT.} Panel for different values of the ratio $\frac{`\nu}{\rho}$ across different values of $N_0$ comparing the different theoretical predictions with the average Heaps Law over $500$ realizations of the process. Each column is relative to a given value of $\frac{`\nu}{\rho}$ while each row relates to a given value of $N_0$. All realizations have length $500000$.}
        \label{fig:heaps_4teorie}
\end{figure}

\begin{figure}[htbp]
        \includegraphics[width=\linewidth]{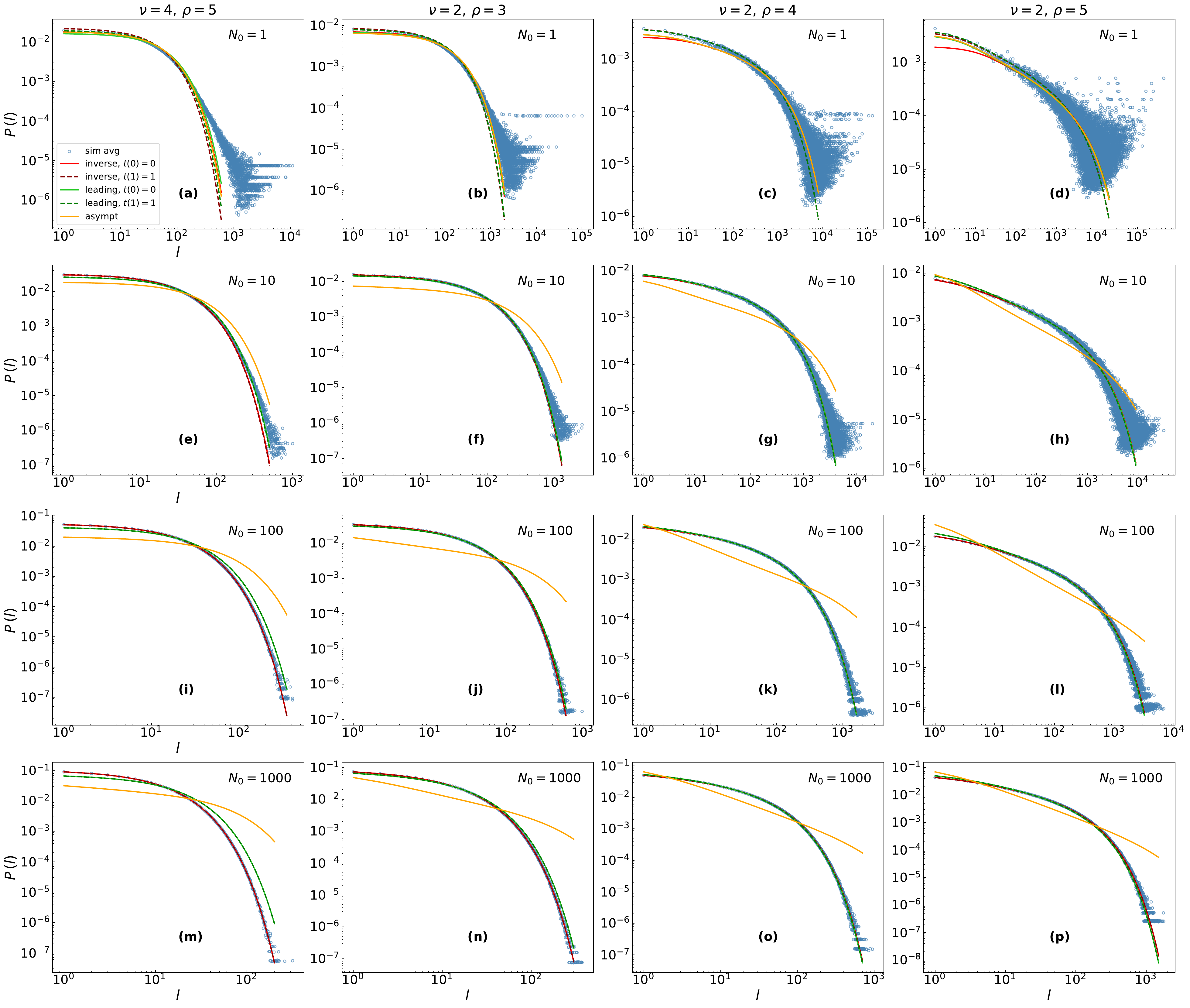}
        \caption{\textbf{Avalanches size distributions for non-novelties in the UMT.} Panel for different values of the ratio $\frac{`\nu}{\rho}$ across different values of $N_0$ comparing the different theoretical predictions with the average distribution over $500$ realizations of the process. As in Fig.~\ref{fig:heaps_4teorie}, each column is relative to a given value of $\frac{`\nu}{\rho}$ while each row relates to a given value of $N_0$. All realizations have length $500000$.}
        \label{fig:delays_4teorie}
\end{figure}

\begin{figure}[htbp]
        \includegraphics[width=0.6\linewidth]{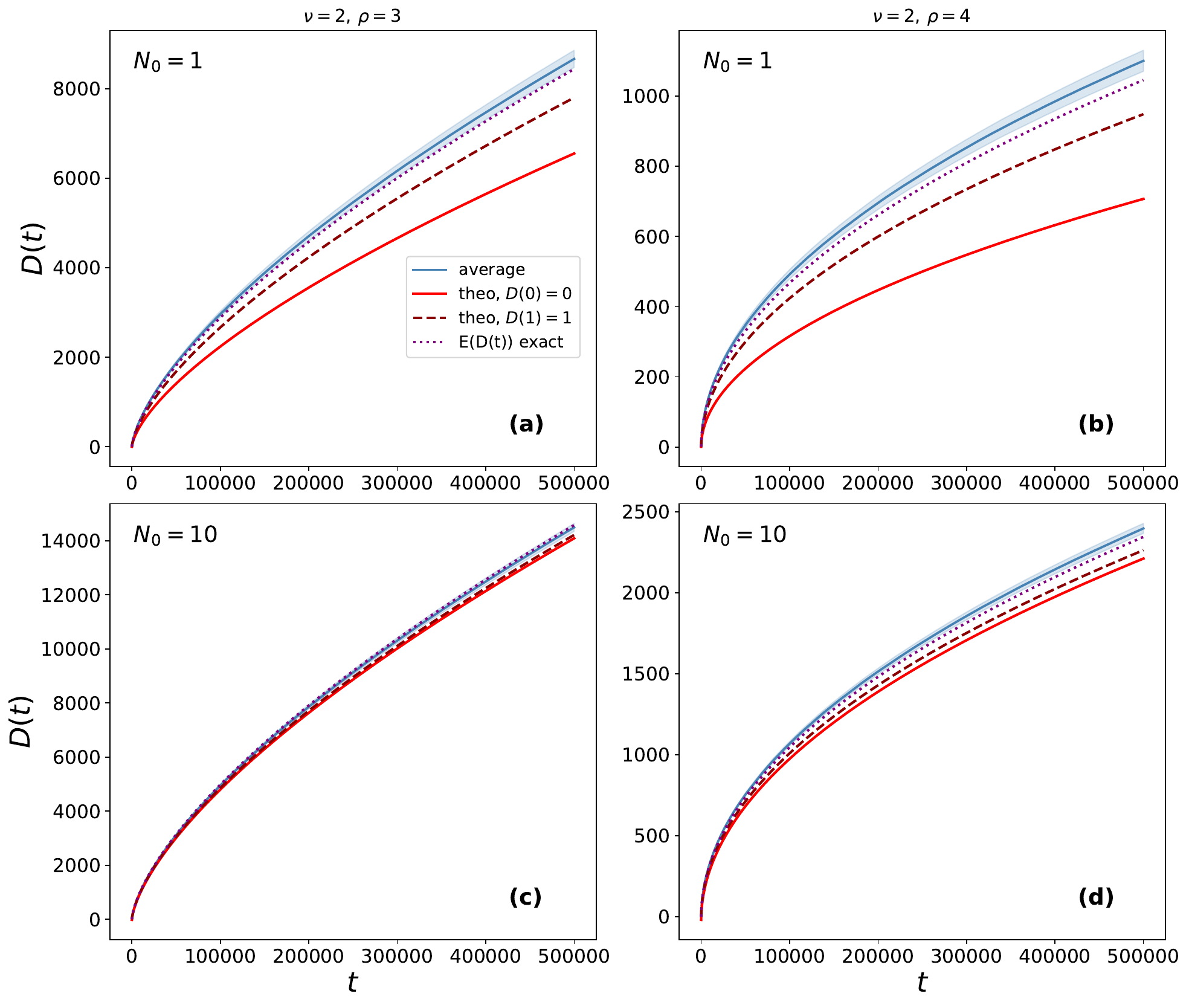}
        \caption{\textbf{Heaps' Law in the UMT-E.} Panel for different values of the ratio $\frac{`\nu}{\rho}$ and for low values of $N_0$, comparing the different theoretical predictions with the average Heaps Law over $500$ realizations of the process.  All realizations have length $500000$.}
        \label{fig:heaps_umt_2x2}
\end{figure}

\begin{figure}[htbp]
        \includegraphics[width=0.6\linewidth]{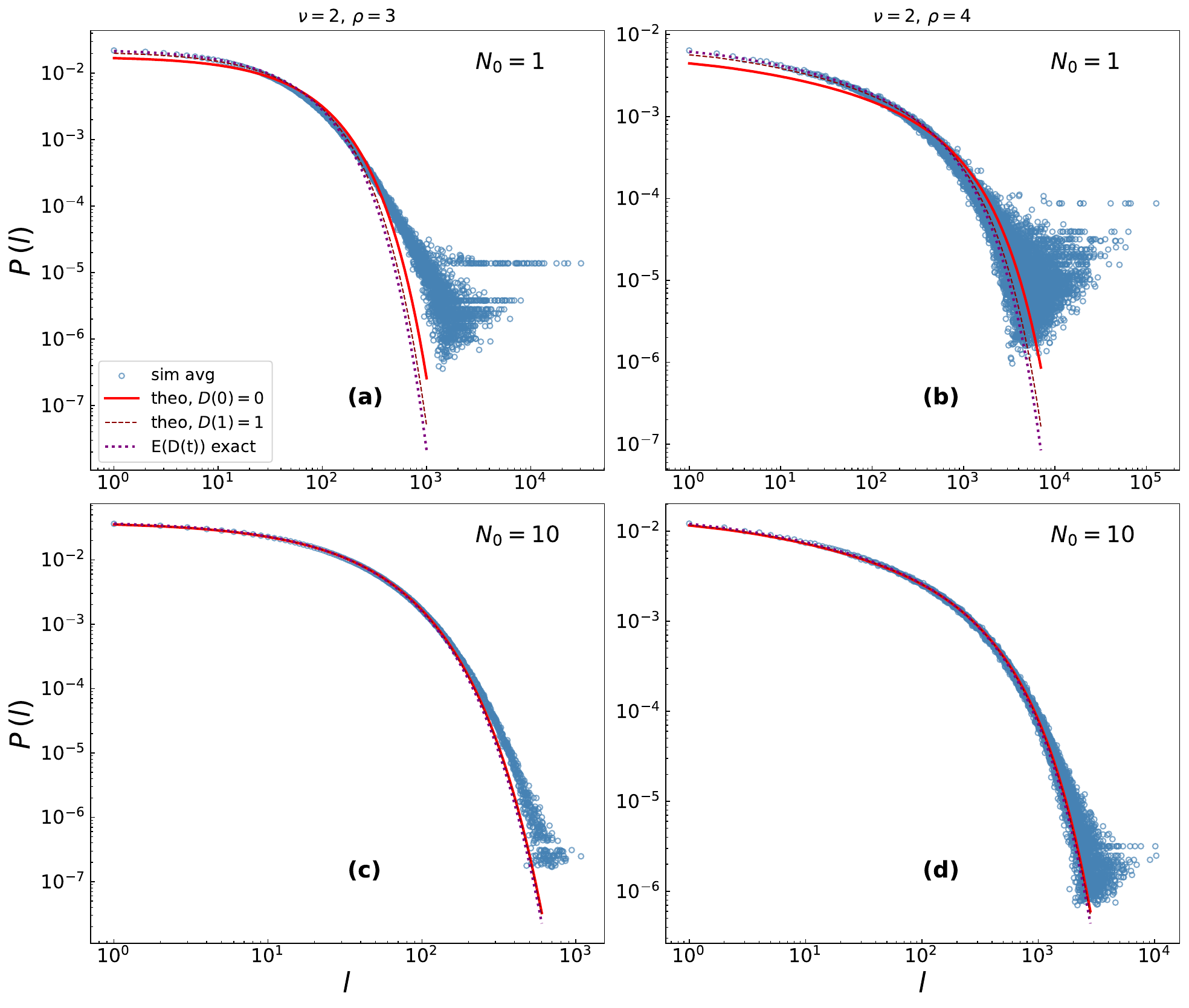}
        \caption{\textbf{Avalanches size distributions for non-novelties in the UMT.} Panel for different values of the ratio $\frac{`\nu}{\rho}$ and for low values of $N_0$, comparing the different theoretical predictions with the average Heaps Law over $500$ realizations of the process.  All realizations have length $500000$.}
        \label{fig:delays_umte_2x2}
\end{figure}

We give here the approximate expressions for $D(t)$ obtained from eq.~\ref{eq:SI_tD} and with initial condition respectively $D(0)=0$ and $D(1)=1$:
\begin{eqnarray}
D_0(t) &= &\frac{N_0}{\nu} \left(\frac{\rho(\rho-\nu)t}{N_0(\rho-\nu+a)} \right)^{\frac{\nu}{\rho}} 
\label{eq:umt_dt_0}
\end{eqnarray}
\begin{eqnarray}
D_1(t) &= &\left(\frac{N_0+\nu}{\nu}\right) \left(\frac{\rho(\rho-\nu)t}{(N_0+\rho)(\rho-\nu+a)} \right)^{\frac{\nu}{\rho}} 
\label{eq:umt_dt_1}
\end{eqnarray}
\noindent
We also report here below the equivalent expressions for the UMT-E model:
\begin{eqnarray}\label{eq:SI_D0-ex}
D_0(t) &= &\frac{N_0}{\nu} \left(1 + \frac{\rho t}{N_0} \right)^{\frac{\nu}{\rho}} - \frac{N_0}{\nu} = \nonumber\\
&=&\frac{\theta}{\alpha} \left(1 + \frac{ t}{\theta} \right)^{\alpha} - \frac{\theta}{\alpha}
\end{eqnarray}
\begin{eqnarray}\label{eq:SI_D1-ex}
D_1(t) &= &\frac{N_0+\nu}{\nu}\left(\frac{N_0+\rho t}{N_0+\rho}\right)^{\frac{\nu}{\rho}} -\frac{N_0}{\nu}= \nonumber\\
&=&\left(\frac{\theta}{\alpha} +1\right) \left(\frac{ \theta + t}{\theta +1} \right)^{\alpha} - \frac{\theta}{\alpha}
\end{eqnarray}
\noindent
In Fig.~\ref{fig:heaps_4teorie} we contrast the predictions for the Heaps law obtained with: i) the inverse relation reported in Eq. (6) in the main text ; ii-iii) the expression reported respectively in Eq.~\ref{eq:umt_dt_0} and~\ref{eq:umt_dt_1}; iv) the asymptotic form reported in (\cite{tria2014dynamics}) and that we report here for completeness:
\begin{eqnarray}
D(t) &= & (\rho-\nu)^{\frac{\nu}{\rho}}t^{\frac{\nu}{\rho}}
\label{eq:umt_asympt}
\end{eqnarray}
In Fig.~\ref{fig:delays_4teorie} we report the results for the avalanche size distributions for non-novelties. We observed that, while for $N_0\simeq 1$ the asymptotic form in~\cite{tria2014dynamics} approaches the results obtained by taking into account the $N_0$ parameter, the former is no longer a good approximation when $N_0 >> 1$, as often happens in real datasets.
\\
In Fig.~\ref{fig:heaps_umt_2x2}, we report a throughout analysis
of the Heaps' law  for the UMT-E model.
We compare the numerical simulation — averaged over 500  independent realizations — against the two analytical solutions in eqs.~\ref{eq:SI_D0-ex} and~\ref{eq:SI_D1-ex}, and an exact form derived in~\cite{zhang2015explicit}, that reads:
\begin{eqnarray}
E[D(t)] &= & \frac{\langle \frac{N_0+\nu}{\rho}\rangle_t}{\langle\frac{N_0}{\rho}\rangle_t} \cdot \frac{N_0}{\nu}-\frac{N_0}{\nu}
\label{eq:umte_exact_heaps}
\end{eqnarray}
where $\langle x\rangle_t= x (x+1) \cdots (x+t-1) $ is the Pochhammer symbol (rising factorial).  When $N_0 \approx 1$, the exact formula yields the closest agreement with the numerical average. However, as long as $N_0$ exceeds a few units, the three theoretical predictions  become indistinguishable within visual resolution. Finally, Fig.~\ref{fig:delays_umte_2x2} presents the corresponding intermission-time distributions for the same parameter values used in Fig.~\ref{fig:heaps_umt_2x2}.

\clearpage
\section{S6. A note on exchangeability}

\begin{figure}[htbp]
        \includegraphics[width=0.5\linewidth]{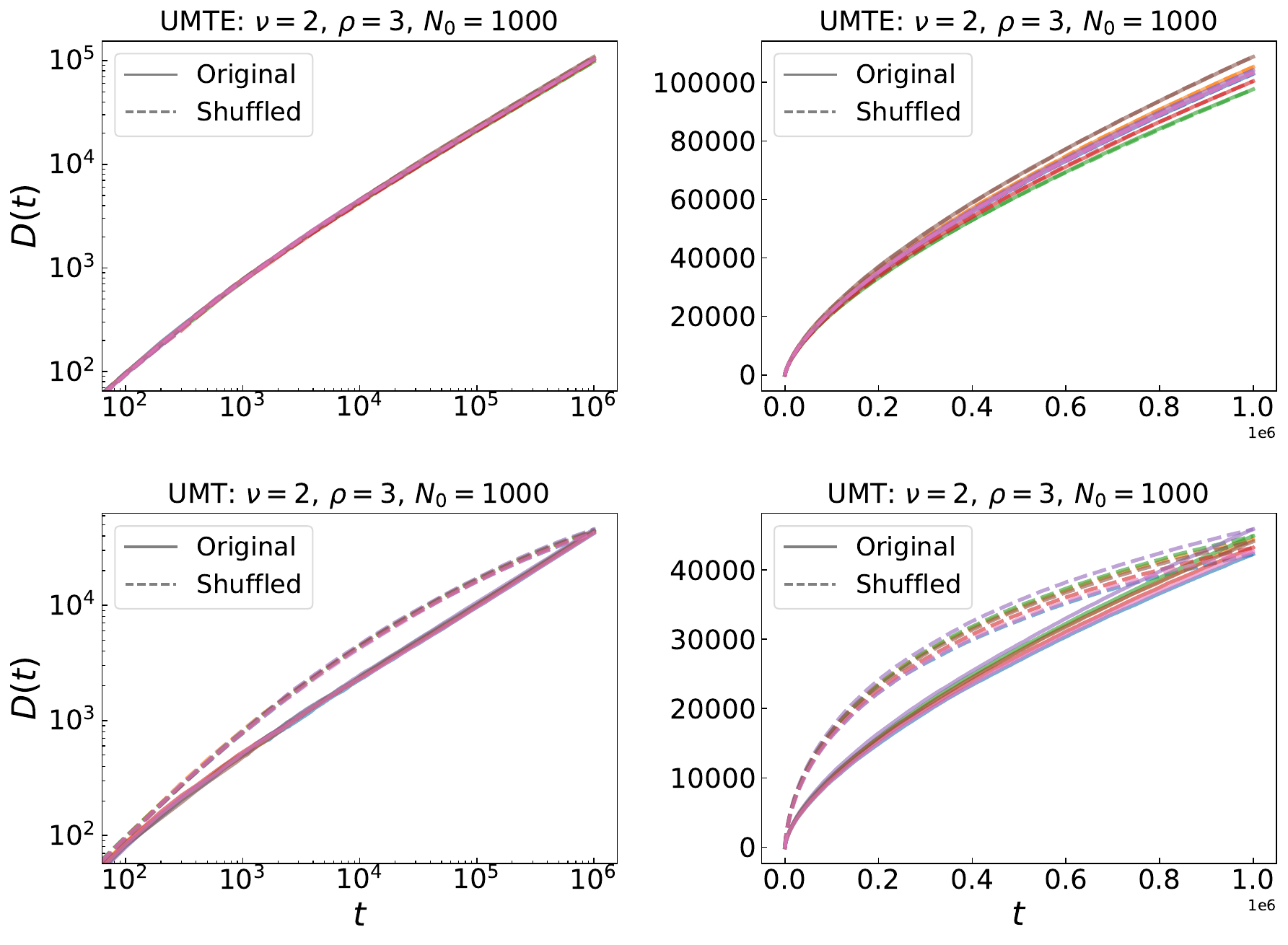}
        \caption{\textbf{Heaps' Law in the UMT and UMT-E sequences.} Comparison between the Heaps Law in the UMT and UMT-E sequences and their corresponding shuffled sequences, for five independent instances in each plot. Left: log scale; right: linear scale; top: UMT-E model; bottom: UMT model.}
        \label{fig:heaps_umt_umte_shuffled}
\end{figure}

\begin{figure}[htbp]
        \includegraphics[width=0.6\linewidth]{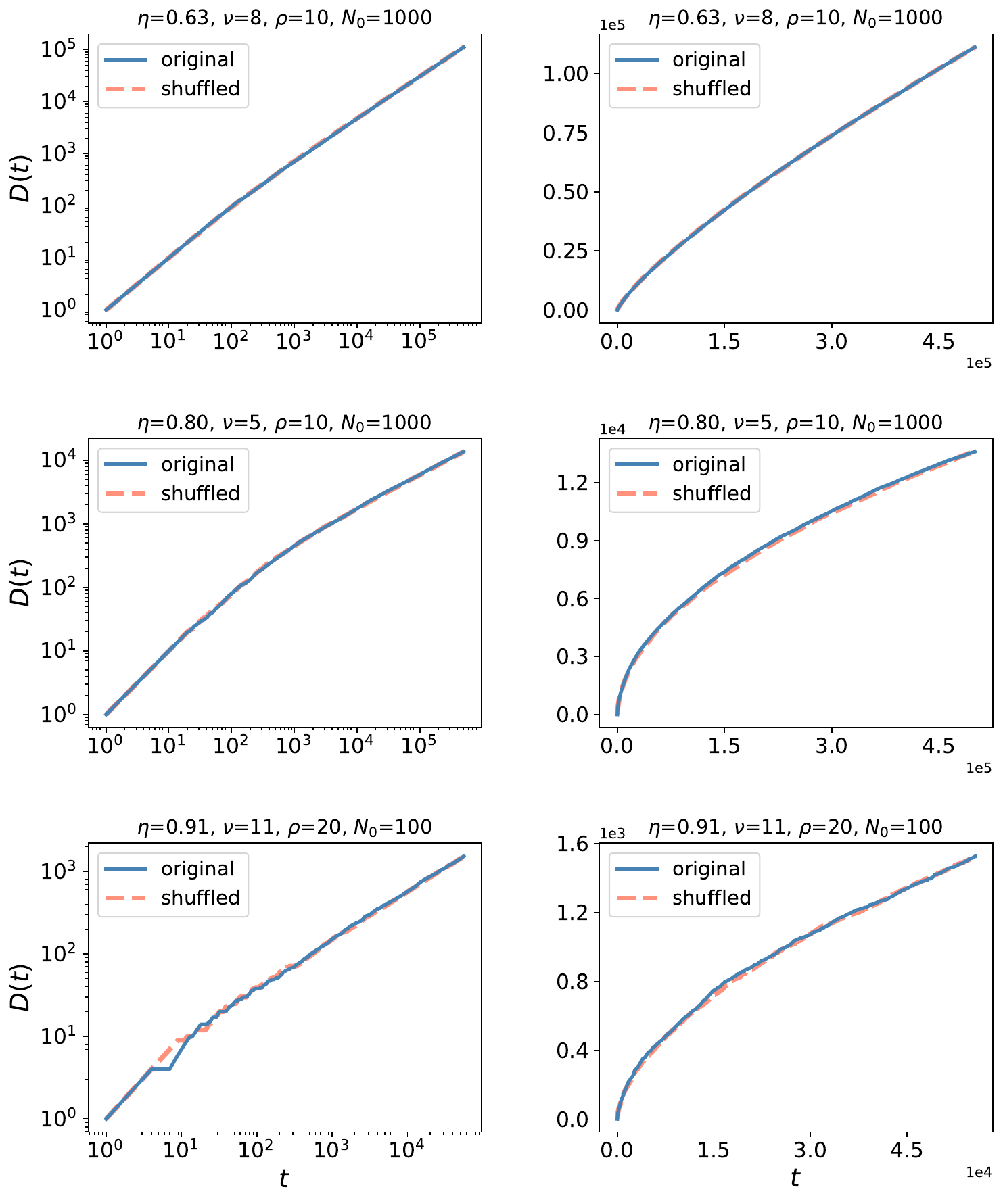}
        \caption{\textbf{Heaps' Law in the UMST-E sequences.} Heaps Law in the UMST-E sequences and their corresponding shuffled sequences. Left: log scale; right: linear scale.}
        \label{fig:heaps_umstexc}
\end{figure}

In Figure~\ref{fig:heaps_umt_umte_shuffled}, we report the results obtained for Heaps’ law in the UMT and UMT-E models and compare them with those obtained from the corresponding shuffled sequences. We show that, as expected, in the exchangeable model the two curves—corresponding to the original and shuffled sequences—overlap, whereas this is not the case for the general UMT model, providing a clear signature of non-exchangeability in the latter. We note that a departure of the Heaps law curve for the shuffled sequence from that of the original sequence is a sufficient but not necessary condition for non-exchangeability. In Figure~\ref{fig:heaps_umstexc}, we report results for the UMST model in which we set $\tilde{\rho}=\rho-(\nu+1)$, so that it reduces to UMT-E when the parameter $\eta$ is set to $1$. In this case, the model’s non-exchangeability has a different origin than in the UMT model, and it is not clearly reflected in Heaps’ law.

\begin{figure}[htbp]
        \includegraphics[width=\linewidth]{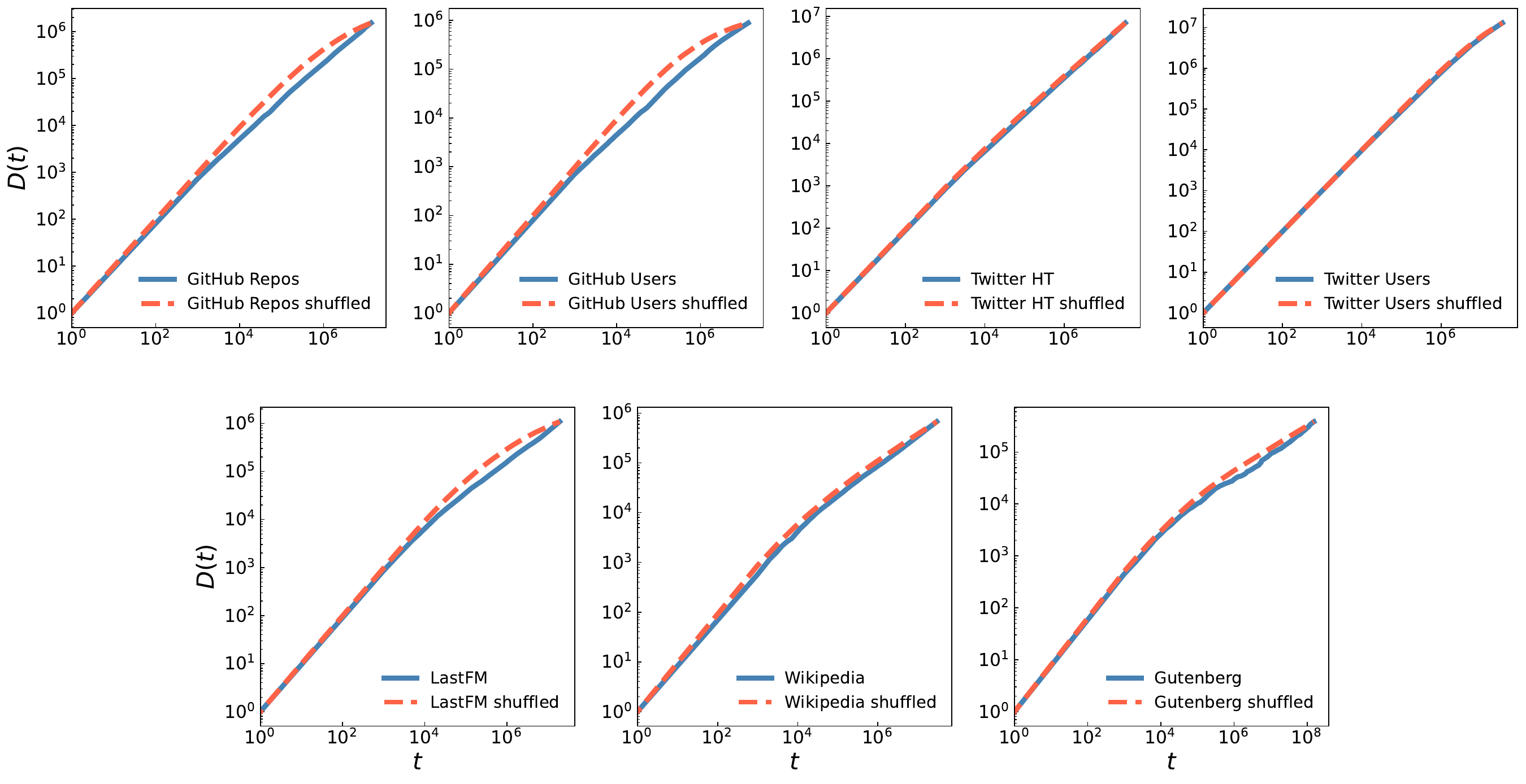}
        \caption{\textbf{Heaps' Laws in real datasets and in their corresponding shuffled sequences (log scale).} Comparison between the Heaps Law in real datasets and their corresponding shuffled sequences.}
        \label{fig:heaps_real_datasets_shuffled}
\end{figure}

\begin{figure}[htbp]
        \includegraphics[width=\linewidth]{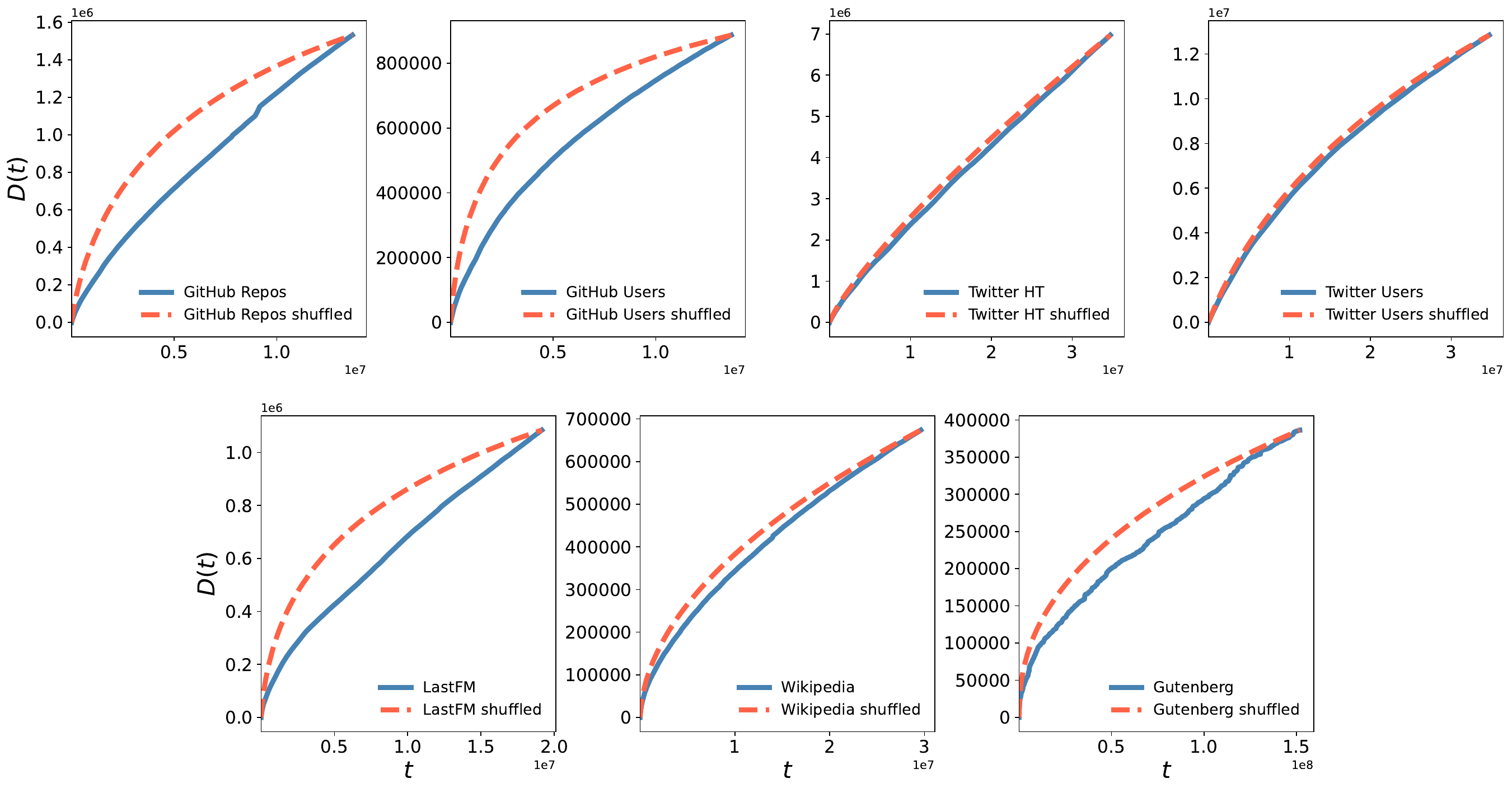}
        \caption{\textbf{Heaps' Laws in real datasets and in their corresponding shuffled sequences (linear scale).} Comparison between the Heaps Law in real datasets and their corresponding shuffled sequences.}
        \label{fig:heaps_real_datasets_shuffled_linear}
\end{figure}

We further contrast Heaps’ law for the original and shuffled sequences across all datasets considered in the main text. These results are reported in Figure~\ref{fig:heaps_real_datasets_shuffled} (log-log scale) and in Figure~\ref{fig:heaps_real_datasets_shuffled_linear} (linear scale). A  recent discussion on this topic can be found   in~\cite{zimmerlin2026dynamics}.
\section{S7. Series expansion of the avalanche size distribution in the UMT-E}
Here, we give an alternative expression for the avalanche size distribution of non-novelties in terms of an explicit series expansion. 

In deriving this, it is convenient
to consider separately the numerator and the denominator in Eq.(10) in the main text. 
The numerator has the form: 
\begin{equation}
\left(1 + \frac{\rho t}{N_0} \right)^{\frac{2 \nu}{\rho} - 2} \left( 1-\left(1 + \frac{\rho t}{N_0} \right)^{\frac{\nu}{\rho} - 1}\right)^l
\end{equation}
\begin{equation}
    (a+b)^{n} =\sum_{k=0}^{n} \binom{n}{k}(a)^{n-k}b^{k}\,,
\end{equation}
so we obtain
\begin{equation}
\left(1 + \frac{\rho t}{N_0} \right)^{\frac{2 \nu}{\rho} - 2} \sum_{s=0}^{L}\binom{L}{s}(-1)^{s} \left(1 + \frac{\rho t}{N_0} \right)^{(\frac{\nu}{\rho} - 1)s}
\end{equation}
\begin{equation}
\sum_{s=0}^{L}\binom{L}{s}(-1)^{s} \left(1 + \frac{\rho t}{N_0} \right)^{(\frac{\nu}{\rho} - 1)(s+2)}
\end{equation}
and by integrating between $t_{max}$ and $0$ we get:
\begin{equation}
\sum_{s=0}^{L}\binom{L}{s}(-1)^{s} \frac{\rho}{(s+2)(\nu-\rho)+\rho} \frac{N_0}{\rho}\left(1 + \frac{\rho t}{N_0} \right)^{(\frac{\nu}{\rho} - 1)(s+2)+1}\big|_{0}^{t_{max}} =
\end{equation}
\begin{equation}
\sum_{s=0}^{L}\binom{L}{s}(-1)^{s} \frac{N_0}{(s+2)(\nu-\rho)+\rho} [ \left(1 + \frac{\rho t_{max}}{N_0} \right)^{(\frac{\nu}{\rho} - 1)(s+2)+1}-1]
\end{equation}
The denominator instead has the form:
\begin{equation}
\left(1 + \frac{\rho t}{N_0} \right)^{\frac{ \nu}{\rho} - 1} \left( 1-\left(1 + \frac{\rho t}{N_0} \right)^{\frac{\nu}{\rho} - 1}\right) =\left(1 + \frac{\rho t}{N_0} \right)^{\frac{ \nu}{\rho} - 1} -\left(1 + \frac{\rho t}{N_0} \right)^{\frac{2\nu}{\rho} - 2}=
\end{equation}
that integrating between $0$ and $t_{max}$ is
\begin{equation}
\frac{N_0}{\nu}[\left(1 + \frac{\rho t_{max}}{N_0} \right)^{\frac{ \nu}{\rho} }-1] -\frac{N_0}{2\nu-\rho}[
\left(1 + \frac{\rho t_{max}}{N_0} \right)^{\frac{2\nu}{\rho} - 1} -1] \simeq \frac{N_0}{\nu}\left(1 + \frac{\rho t_{max}}{N_0} \right)^{\frac{ \nu}{\rho} }
\end{equation}
By putting everything together, we get:
\begin{equation}
P(L)\simeq \sum_{s=0}^{L}\binom{L}{s}(-1)^{s} \frac{\nu}{(s+2)(\nu-\rho)+\rho}  \left(1 + \frac{\rho t_{max}}{N_0} \right)^{(\frac{\nu}{\rho} - 1)(s+1)}
\end{equation}
\section{S8. Equivalence between the UMT-E, UMT, and UMST with respect to the avalanche statistics}
In this section, we demonstrate the equivalence of the UMT-E model with the general UMT, the UMST and the UMST model in its exchangeable version in terms of Heaps' law and avalanches of non-novelty distributions. We show that it is always possible to identify a triple $(\nu.\rho,N_{0})$ for the UMT-E model  that reproduces the statistical behavior observed in the other models. The fitting procedure follows the same methodology adopted for real-world datasets. Given a realization from a specific model, we fit its Heaps' law using the analytical UMT-E Heaps law expression and estimate the corresponding parameters $(\nu,\rho,N_0)$. Subsequently, we compute the theoretical avalanche size distribution of non-novelties.

\begin{figure}[htbp]
        \includegraphics[width=\linewidth]{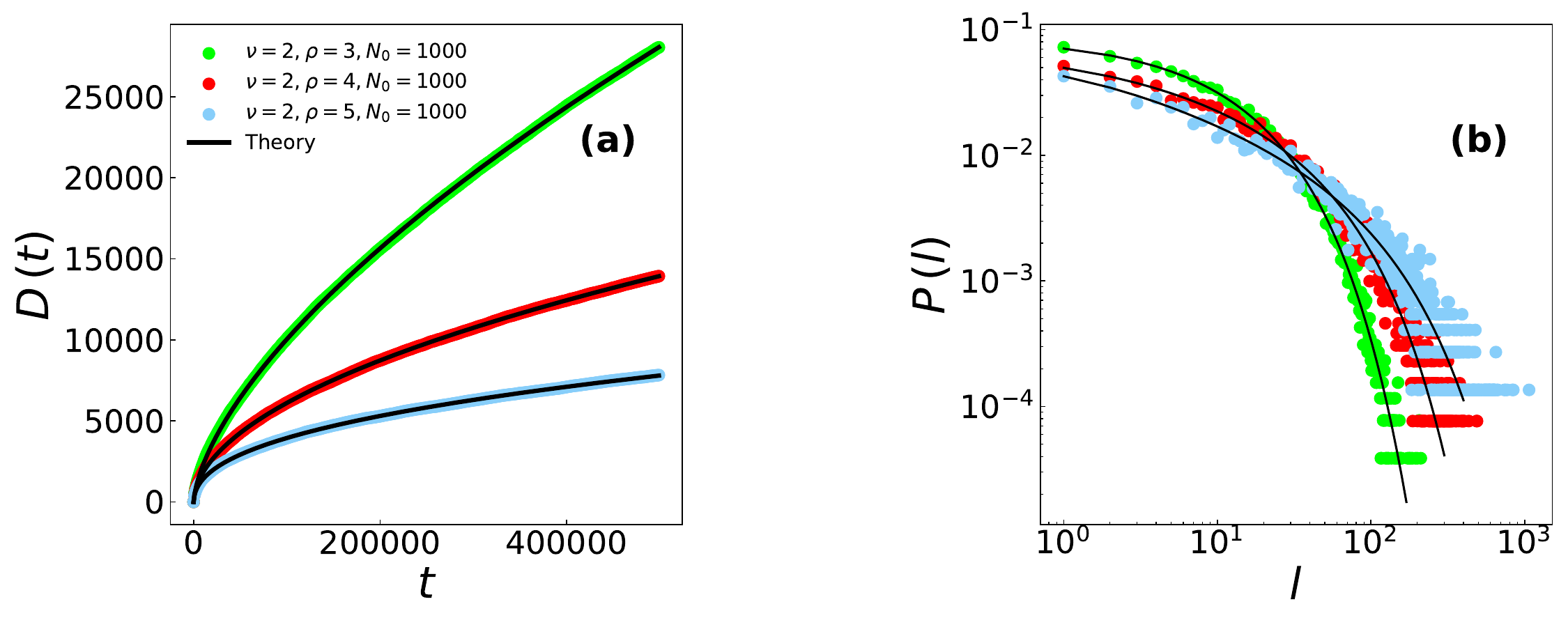}
        \caption{\textbf{Equivalence between the UMT-E and the UMT.} \textbf{a)} Heaps' law for three UMT model realizations and corresponding analytical UMT-E Heaps law predictions. The UMT parameter sets $(\nu=2,\rho=3,N_{0}=1000)$,$(\nu=2,\rho=4,N_{0}=1000)$,$(\nu=2,\rho=5,N_{0}=1000)$ are contrasted with the UMT-E parameter sets $(\nu=7,\rho=11,N_{0}=583)$,$(\nu=1,\rho=2,N_{0}=194)$,$(\nu=21,\rho=52,N_{0}=5593)$, respectively.  \textbf{b)} Avalanche size distributions for non-novelties of the UMT realizations, along with the theoretical UMT-E avalanches size distributions for non-novelties. All the sequences have length $5\times 10^{5}$ and we use the same parameter values as in plot a).}
        \label{fig:umte_spiega_umt}
\end{figure}

\begin{figure}[htbp]
        \includegraphics[width=\linewidth]{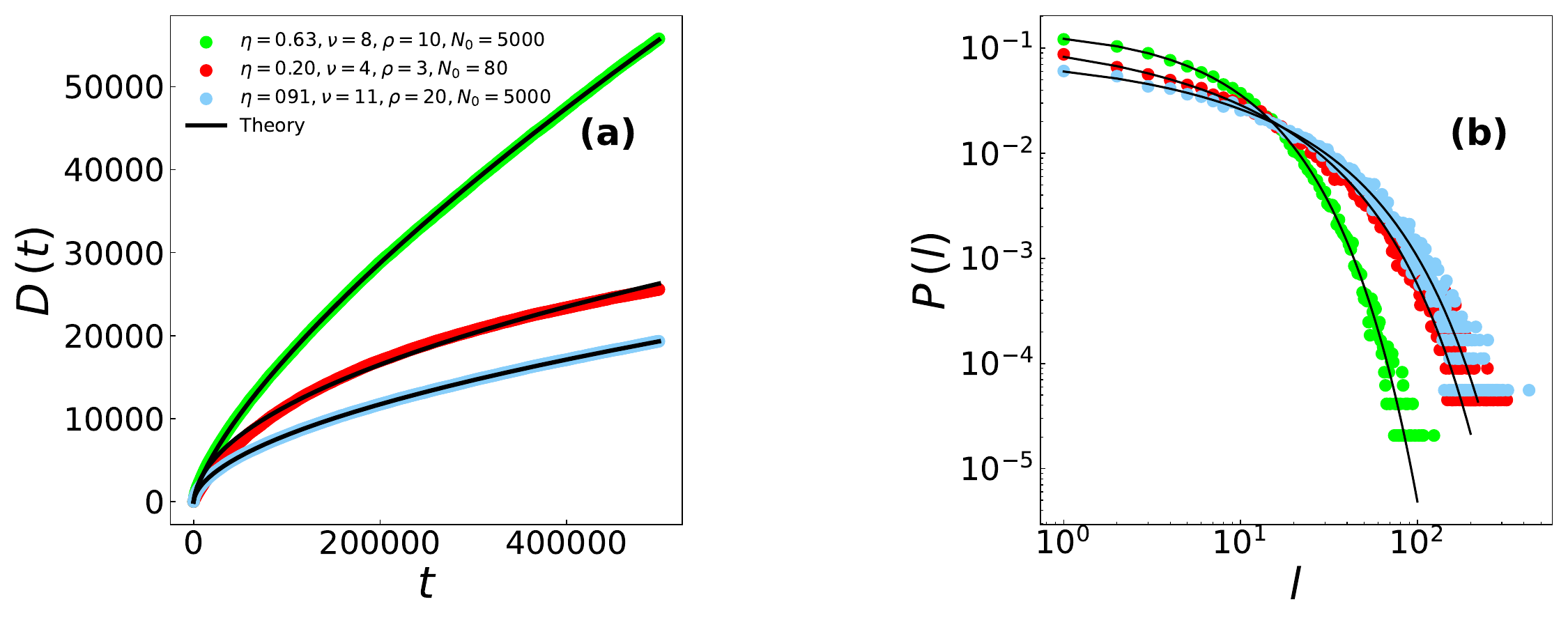}
        \caption{\textbf{Equivalence between the UMT-E and the UMST.} \textbf{a)} Heaps' law for three UMST model realizations, and the corresponding analytical UMT-E Heaps law predictions. The UMST parameter sets $(\eta=0.63,\nu=8,\rho=10,N_{0}=500)$,$(\eta=0.20,\nu=4,\rho=30,N_{0}=80)$,$(\eta=0.91,\nu=11,\rho=20,N_{0}=1000)$ are contrasted with the UMT-E parameter sets $(\nu=31,\rho=43,N_{0}=2570)$,$(\nu=22,\rho=45,N_{0}=28509)$,$(\nu=50,\rho=93,N_{0}=10937)$, respectively .  \textbf{b)} Avalanche size distributions for non-novelties of the UMST realizations, along with the theoretical UMT-E avalanches size distributions for non-novelties. All the sequences have length $5\times 10^{5}$ and we use the same parameter values as in plot a).}
        \label{fig:umte_spiega_umst}
\end{figure}

\begin{figure}[htbp]
        \includegraphics[width=\linewidth]{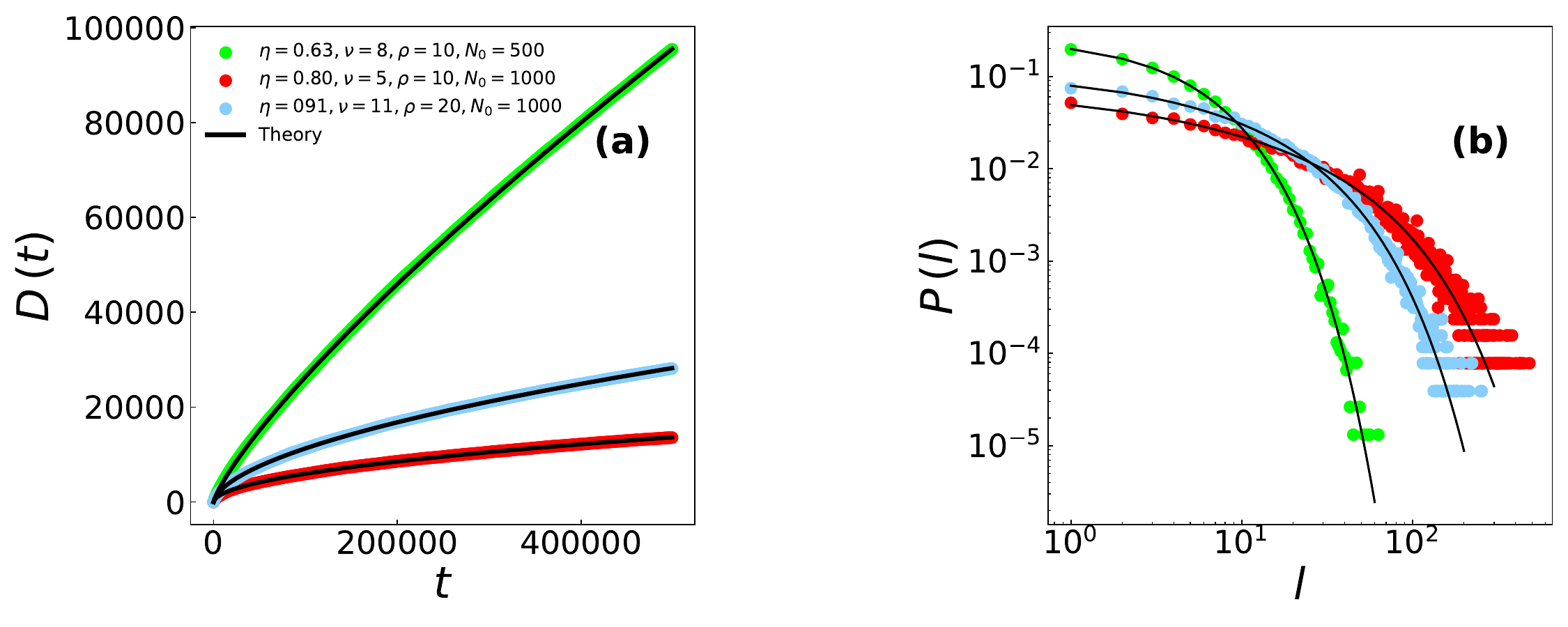}
        \caption{\textbf{Equivalence between UMT-E and UMST-E models.} 
\textbf{a)} Heaps' law for three UMST-E model realizations along with the corresponding analytical UMT-E Heaps Law predictions. The UMST-E parameter sets 
$(\eta=0.63, \nu=8, \rho=10, N_0=500)$, 
$(\eta=0.80, \nu=5, \rho=10, N_0=1000)$, and 
$(\eta=0.91, \nu=11, \rho=20, N_0=1000)$ 
are contrasted with the UMT-E parameter sets 
$(\nu=79, \rho=99, N_0=4346)$, 
$(\nu=50, \rho=99, N_0=9128)$, and 
$(\nu=29, \rho=52, N_0=10804)$, respectively. 
\textbf{b)} Avalanche size distributions for non-novelties of the UMT realizations, along with the theoretical UMT-E avalanche size distributions for non-novelties. 
All the sequences have length $5 \times 10^5$ and we use the same parameter values as in plot a).}
        \label{fig:umte_spiega_umste}

\end{figure}

Figure~\ref{fig:umte_spiega_umt} demonstrates the equivalence between the UMT-E and UMT model (in the $\nu < \rho$ regime). Figure~\ref{fig:umte_spiega_umst} shows the equivalence between the UMT-E and UMST model. In Figure~\ref{fig:umte_spiega_umste}, we additionally present the UMST-E variant, which corresponds to the UMST model with $\tilde{\rho} = \rho - (\nu+1)$. This variant is an analogue of the UMT-E but lacks the exchangeability property.
\\
In tables (\ref{tab:umt_parameters},\ref{tab:umst_parameters},\ref{tab:umst_e_parameters}) we resume the equivalent parameters between each model. 
\begin{table}[htbp]
\centering
\begin{tabular}{cc@{\hspace{0.3cm}}c|cc@{\hspace{0.3cm}}c}
\toprule
\multicolumn{3}{c|}{\textbf{UMT Parameters}} & \multicolumn{3}{c}{\textbf{Equivalent UMT-E Parameters}} \\
\midrule
$\nu$ & $\rho$ & $N_0$ & $\nu$ & $\rho$ & $N_0$ \\
\midrule
2 & 3 & 1000 & 7 & 11 & 583 \\
2 & 4 & 1000 & 1 & 2 & 194 \\
2 & 5 & 1000 & 21 & 52 & 5593 \\
\bottomrule
\end{tabular}
\caption{Equivalent parameter sets between UMT and UMT-E models}
\label{tab:umt_parameters}
\end{table}

\begin{table}[htbp]
\centering
\begin{tabular}{ccc@{\hspace{0.3cm}}c|cc@{\hspace{0.3cm}}c}
\toprule
\multicolumn{4}{c|}{\textbf{UMST Parameters}} & \multicolumn{3}{c}{\textbf{Equivalent UMT-E Parameters}} \\
\midrule
$\eta$ & $\nu$ & $\rho$ & $N_0$ & $\nu$ & $\rho$ & $N_0$ \\
\midrule
0.63 & 8 & 10 & 500 & 31 & 43 & 2570 \\
0.20 & 4 & 3 & 80 & 22 & 45 & 28509 \\
0.91 & 11 & 20 & 1000 & 50 & 93 & 10937 \\
\bottomrule
\end{tabular}
\caption{Equivalent parameter sets between the UMST model and the UMT-E model.}
\label{tab:umst_parameters}
\end{table}

\begin{table}[htbp]
\centering
\begin{tabular}{ccc@{\hspace{0.1cm}}c|ccc}
\toprule
\multicolumn{4}{c|}{\textbf{UMST-E parameters}} & \multicolumn{3}{c}{\textbf{Equivalent UMT-E Parameters}} \\
\midrule
$\eta$ & $\nu$ & $\rho$ & $N_0$ & $\nu$ & $\rho$ & $N_0$ \\
\midrule
0.63 & 8 & 10 & 500 & 79 & 99 & 4346 \\
0.80 & 5 & 10 & 1000 & 50 & 99 & 9128 \\
0.91 & 11 & 20 & 1000 & 29 & 52 & 10804 \\
\bottomrule
\end{tabular}
\caption{Equivalent parameter sets between the UMST-E model and the UMT-E model.}
\label{tab:umst_e_parameters}
\end{table}

\end{document}